\documentclass[preprint]{elsarticle}
\usepackage{textcomp}
\usepackage{amsmath}

\def\be{\begin{equation}}
\def\ee{\end{equation}}
\def\bi{\begin{itemize}}
\def\ei{\end{itemize}}
\def\de{$^{\circ}$C}
\def\tilde{\raisebox{-0.25\baselineskip}{\textasciitilde}}

\newcommand{\sub}[1]{\ensuremath{_{\text{{#1}}}}}

\begin{document}

\begin{frontmatter}

\title{The flight of the GAPS prototype experiment}

\author[ucb,uh]{P.~von Doetinchem\corref{cor1}}
\ead{philipvd@hawaii.edu}
\author[cu]{T.~Aramaki}
\author[jax]{N.~Bando}
\author[ucb]{S.E.~Boggs}
\author[jax]{H.~Fuke}
\author[cu]{F.H.~Gahbauer}
\author[cu]{C.J.~Hailey}
\author[cu]{J.E.~Koglin}
\author[ucla]{S.A.I.~Mognet}
\author[cu]{N.~Madden}
\author[jax]{S.~Okazaki}
\author[ucla]{R.A.~Ong}
\author[cu]{K.M.~Perez}
\author[jax]{T.~Yoshida}
\author[ucla]{J.~Zweerink}

\cortext[cor1]{Corresponding author}

\address[ucb]{Space Sciences Laboratory, University of California, Berkeley, USA}
\address[uh]{Department of Physics and Astronomy, University of Hawai'i at M$\bar{\text{a}}$noa, Honolulu, USA}
\address[cu]{Department of Physics, Columbia University, New York, USA}
\address[jax]{Institute of Space and Astronautical Science, Japan Aerospace Exploration Agency (ISAS/JAXA), Sagamihara, Japan}
\address[ucla]{Department of Physics and Astronomy, University of California, Los Angeles, USA}

\begin{abstract}
The General AntiParticle Spectrometer experiment (GAPS) is foreseen to carry out a dark matter search using low-energy cosmic ray antideuterons at stratospheric altitudes with a novel detection approach. A prototype flight from Taiki, Japan was carried out in June 2012 to prove the performance of the GAPS instrument subsystems (Lithium-drifted Silicon tracker and time-of-flight) and the thermal cooling concept as well as to measure background levels. The flight was a success and the stable flight operation of the GAPS detector concept was proven. During the flight about $10^6$ charged particle triggers were recorded, extensive X-ray calibrations of the individual tracker modules were performed by using an onboard X-ray tube, and the background level of atmospheric and cosmic X-rays was measured. The behavior of the tracker performance as a function of temperature was investigated. The tracks of charged particle events were reconstructed and used to study the tracking resolution, the detection efficiency of the tracker, and coherent X-ray backgrounds. A timing calibration of the time-of-flight subsystem was performed to measure the particle velocity. The flux as a function of flight altitude and as a function of velocity was extracted taking into account systematic instrumental effects. The developed analysis techniques will form the basis for future flights.
\end{abstract}


\end{frontmatter}

\section{Introduction}

\subsection{Indirect dark matter search with antideuterons}

The existence of dark matter and its nature play a key role in understanding structure formation after the big bang and the energy density of the universe \cite{darkmatter}. Dark matter cannot be explained with known types of matter; therefore, we are at the dawn of something significantly new. The importance of this problem becomes obvious by recalling that dark matter is approximately five times more abundant than regular matter. Little is known about the nature of dark matter particles, but that they are relatively heavy, interacting gravitationally with regular matter, but otherwise, interacting only weakly if at all. If dark matter was in thermal equilibrium with the rest of the matter in the early universe and froze out when the temperature dropped due to expansion, it is a natural assumption in many models that dark matter particles are able to annihilate with each other and produce known standard model particles in this way. These particles would contribute to the known cosmic ray fluxes and, as the kinematic characteristics of these processes are different from the production mechanisms of the conventional cosmic rays, it could be possible to observe the imprint of dark matter in the diffuse cosmic ray spectra in the form of an excess. Well-motivated theories beyond the standard model of particle physics contain candidates with exactly these properties. Cosmic ray antiparticles -- without primary sources -- are ideal candidates for such a search. However recent results show that accomplishing this task with positrons and antiprotons appears to be difficult \cite{pamela,pbarpamela,amsposi1,amsposi2}.

Antideuterons would also be generated in dark matter annihilations and are a potential breakthrough approach. Secondary antideuterons, like antiprotons, are produced when cosmic ray protons or antiprotons interact with the interstellar medium, but the production threshold for this reaction is higher for antideuterons than antiprotons. Collision kinematics also disfavor the formation of low-energy antideuterons in these interactions. Moreover the steep energy spectrum of cosmic rays means there are fewer particles with sufficient energy to produce secondary antideuterons, and those that are produced will have relatively large kinetic energy. As a consequence, a low-energy search for primary antideuterons has very low background \cite{dbarsusy,dbarprod,ibarra20132,forn}. This feature of antideuteron searches, along with the growing realization that such searches probe supersymmetric models in a broad way, has attracted considerable attention. Many theoretical papers discuss aspects of antideuteron dark matter searches \cite{dbarpbh,dbarpbh2,dbarbaer,dbarback,kadastik2010,antideuteroncui,gravi,ibarra20131}. In this regard, supersymmetric and universal extra dimension theories provide the most popular and theoretically well-motivated dark matter candidates \cite{susydm,kkdm}. The absolute flux expected for antideuterons is very low, and therefore any attempt to measure it needs an exceptionally strong particle identification.

\subsection{Detection of antideuterons in cosmic rays}

\begin{figure}
\centerline{\includegraphics[width=0.7\linewidth]{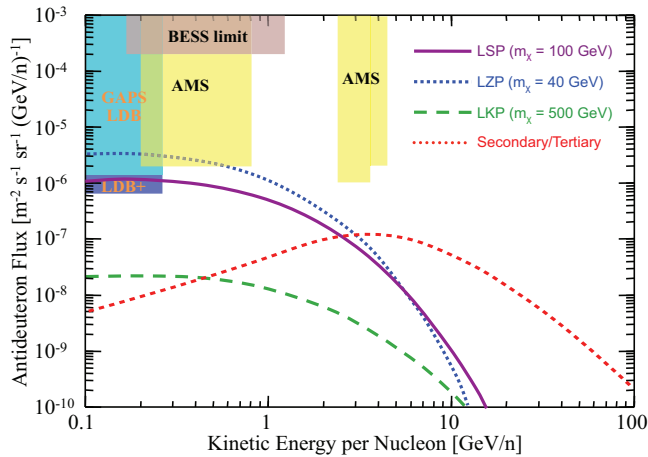}}
\caption{\label{f-fig1_jap_pgaps}Predicted antideuteron fluxes from different dark matter models updated by more recent coalescence momentum value (purple, red, green lines) \cite{dbarbaer,ibarra20131} and secondary/tertiary background flux from cosmic ray interactions with the interstellar medium (blue line) \cite{ibarra20132}. Antideuteron limits from BESS \cite{bess} and sensitivities for the running AMS \cite{dbarsens} and the planned GAPS experiments \cite{gapssens} are also shown.}
\end{figure}

In the near future the challenging search for antideuterons will exclusively rely on the General AntiParticle Spectrometer (Section~\ref{s-gaps}) and on the Alpha Magnetic Spectrometer (AMS), a multi-purpose cosmic ray detector on the International Space Station \cite{gaps,amsdbar}. Figure~\ref{f-fig1_jap_pgaps} shows the theoretically expected antideuteron fluxes from different dark matter models in comparison to the secondary background. The different boxes demonstrate the antideuteron flux limits of BESS \cite{bess} and the sensitivity reaches of GAPS and AMS \cite{gapssens,dbarsens}, which reach for the first time the sensitivity to probe predictions of well-motivated models. Both experiments have mostly complementary kinetic energy ranges but also some overlap in the interesting low-energy region. In addition, another very important virtue comes from the different detection techniques. AMS identifies particles by analyzing the event signatures of different subsequent subdetectors and a strong magnetic field and GAPS by slowing down the antideuteron and creating an exotic atom inside the target material and analyzing the decay. This allows the study of both a large energy range and independent experimental confirmation, which is crucial for a rare event search like the hunt for cosmic ray antideuterons.

\subsection{The GAPS experiment\label{s-gaps}}

\subsubsection{Mission overview}

\begin{figure}
\centerline{\includegraphics[width=0.7\linewidth]{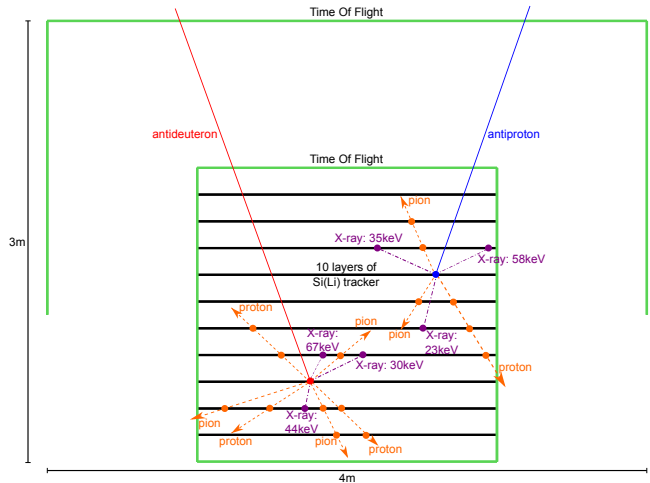}}
\caption{\label{f-fig2_jap_pgaps}GAPS detector concept with antiproton and antideuteron signatures.}
\end{figure}

The General AntiParticle Spectrometer is designed to measure low-energy cosmic antideuterons. As mentioned above, the expected antideuteron flux is very low and therefore a large acceptance and long flight time are indispensable. Figure~\ref{f-fig1_jap_pgaps} demonstrates that the dark matter signal above the background is expected to be the largest at low kinetic energies of 100--500\,MeV. It is therefore planned for GAPS to carry out a series of long duration balloon flights from Antarctica,  where the deflection of low-energy charged cosmic rays in the geomagnetic field is the smallest. Another important effect for the low-energy cosmic ray detection comes from the interaction of the solar wind with cosmic rays, which effectively decreases the observable interstellar flux. The strength of the solar modulation depends on the 11 year long solar cycle, which will approximately reach the next minimum in 2019 \cite{solarcycle}. The first science GAPS flight will be possible from 2017 and will therefore only feel a small solar influence.

\subsubsection{Detector and identification concept}

The core of the detector will be a track reconstruction device consisting of 10 layers of Lithium-drifted Silicon (Si(Li)) modules (Figure~\ref{f-fig2_jap_pgaps}) that will be enclosed by a hermetically sealed time-of-flight system (TOF) made of plastic scintillators with photomultiplier tube (PMT) readout.  This box will be surrounded by another half-cube of plastic scintillators. The inner tracker core (TRK) will be a cube of 2\,m edge length and the outer TOF half-cube will have a width of 4\,m.

These detector components will be used for a novel detection approach to clearly identify low-energy antideuterons. The idea is to stop low-energy antideuterons in the tracker material, to replace a shell electron of the target material with this antideuteron, and to form an excited exotic atom. The Hydrogen-like atom will deexcite by autoionization followed by characteristic X-ray ladder transitions. At the end of the ladder transitions the antideuteron will annihilate in a hadronic interaction with the nucleus and produce pions and protons. The detector will be able to measure the velocity and the charge of the incoming particle in the TOF as well as the stopping depth of a particle in the tracker and the development of the energy loss per layer throughout the slowing process. Moreover, the tracker will resolve the characteristic X-ray energies and track the pions and protons. The main source of background for the antideuteron signal comes from antiprotons. Therefore, good depth sensing and X-ray energy resolution along with a reliable tracking and counting of pions/protons are essential for a high background rejection.

\section{The GAPS prototype experiment}

\begin{figure}
\centerline{\includegraphics[width=0.7\linewidth]{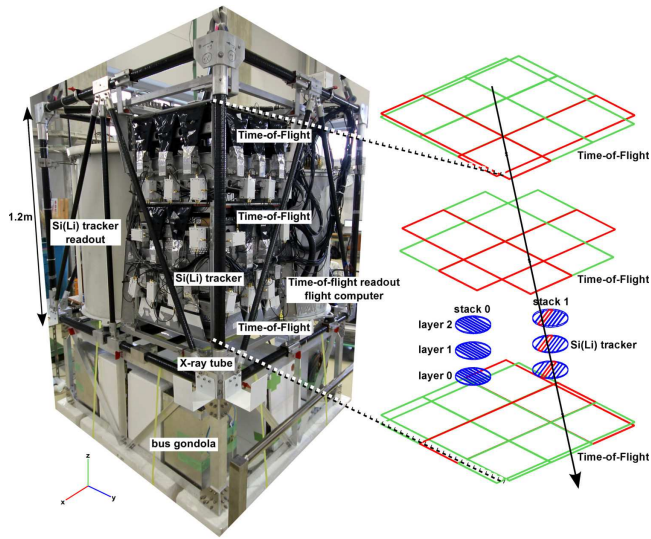}}
\caption{\label{f-fig3_jap_pgaps}\textbf{Left:} Flight ready pGAPS payload without insulation foam. \textbf{Right:} Event display showing the position of the Time-of-Flight and Si(Li) tracker subsystems and a typical clean cosmic ray track reconstructed from flight data.}
\end{figure}

A GAPS prototype (pGAPS) was constructed for a balloon flight to demonstrate a stable and low noise operation of all relevant detector components, to verify the thermal model and the Si(Li) detector module cooling approach, and to study the incoherent background level of charged cosmic rays and X-rays. The small acceptance and short flight time compared to the full-scale experiment did not allow a study of antiprotons or antideuterons. However, the prototype consisted of all major components that will be part of the full GAPS experiment (Figure~\ref{f-fig3_jap_pgaps}). The carbon fiber structure of the balloon gondola frame had a height of 1.2\,m, the mass of the science part of the gondola was 308\,kg, and the total power consumption was 430\,W. A full description of the instrument can be found here \cite{isaacpaper}.

\subsection{Si(Li) tracker}

In total six circular modules were arranged in two stacks of three with a vertical spacing of 20\,cm and mounted into a watertight plastic vessel. The individual circular Si(Li) modules were manufactured by Semikon Detektor GmbH of J\"{u}lich, Germany. For the full payload the detectors will be manufactured by the GAPS collaboration \cite{detdev}. Five modules had a thickness of 2.5\,mm and one of 4.0\,mm. The active area of each module had a diameter of 9.4\,cm and was divided into eight strips. The strips on the p+ side were contacted by implanted Boron and on the n+ side by Lithium contacts. For full depletion the detectors had to be cooled down to at least -25\de~and were operated with reverse bias voltages of 185-240V. The cooling system used a closed-loop coolant pipe with Fluorinert fluid to transport the heat from the detectors to a radiator using a pump. An attitude control system was designed to point the radiator to the anti-Sun side during the flight. On the ground, the radiator was thermally coupled to a heat exchanger cooled by liquid Nitrogen and the detectors were operated in a dry Nitrogen atmosphere.

For the science payload it will be important to resolve X-rays in the range of 10--100\,keV while simultaneously measuring charged energy depositions up to 50\,MeV. Therefore, the pGAPS Si(Li) electronics had a dual readout scheme, with separate high and low gain processing chains for each detector strip, which consisted of two card cages with three analog readout boards each. The card cages, boards, and digital signal processing units had been flown on the Nuclear Compton Telescope prototype in 2005 \cite{nct05} and were modified to meet the requirements for pGAPS. The electronics were designed to work at low pressures and temperatures and were housed in a watertight plastic vessel. The electronics also recorded various scalers important for monitoring the quality of the signal processing of the different detector strips, temperatures, electronic status, and livetime.

For in-flight calibration the payload was equipped with a Silver target X-ray tube and provided, together with a Gold filter, a characteristic spectrum with peaks at 26 and 36\,keV. The tube was mounted under the bottom time-of-flight layer and the position was optimized to illuminate the detectors as uniformly as possible. In addition, the 59.5\,keV line of an Americium-241 (Am-241) radioactive source was used for ground calibration during the qualification, integration, and flight preparation stages. 

\subsection{Time-of-flight detector}

The time-of-flight system (TOF) consisted of three layers of crossed plastic scintillator paddles, two above and one below the tracker. The top and the bottom layers were composed of two individual layers with three paddles each and the middle layer was composed of two layers of two paddles each. The paddles were made of Bicron BC-408 material and were 50\,cm long, 15\,cm wide, and 3\,mm thick. Each paddle end was attached to a curved, acrylic light guide coupled to a Hamamatsu R-7600 photomultiplier tube, which was operated at about 800-900\,V. Prior to the flight, all flight (and flight spare) TOF PMT assemblies were operated at full HV for a minimum of 5\,hours each in a low pressure environment. The test was not a full thermal-vacuum test since the temperature was not controlled. However, the pressure was varied over a range of pressures expected in flight (1--50\,torr), with most of the test taking place at pressures between 5 and 10\,torr. No failures or degraded performance were observed for any PMTs 
in the tests.

The spacing between the top and bottom layer was 0.94\,m and the spacing between the top and middle layer was 0.38\,m. The TOF data were processed in a VME rack with modules. The VME system also made the TOF trigger decision that will be further explained in the next section. In addition to the trigger decision, the electronics digitized the scintillator light output as measured by the PMTs and generated time-to-digital converter (TDC) values with 50\,ps resolution based on a discriminator threshold. Housekeeping data were also recorded. The TOF electronics, flight computer, attitude control system electronics, and a fiber optic gyroscope were mounted together in a pressurized vessel since the TOF electronics modules were not designed to operate in vacuum.

\subsection{Trigger modes\label{s-trig}}

pGAPS was operated in several different data taking modes. The goal was both to collect tracks of charged particles traversing the TOF and the tracker and to measure the stability of the X-ray performance of the tracker modules. Therefore two different trigger schemes were deployed. The TOF trigger mode was based on the coincident signals of two crossed paddles of the TOF middle layer and at least one additional signal in the top or bottom TOF layer, which triggered the simultaneous readout of all PMTs and also of all tracker modules. In this way particle tracks with coincident signals in different detector layers could be recorded.

During the tracker trigger mode each Si(Li) module was able to self-trigger its own readout based on discriminator thresholds. Eventually it will be crucial for GAPS to detect X-rays in coincidence with charged particle tracks, but for calibration purposes no coincidence with other tracker modules or the TOF was required. During flight, the tracker trigger mode was used for X-ray tube calibration and for incoherent X-ray flux measurements. In the tracker trigger mode the TOF continued regular data taking as described above, but without triggering the readout of the tracker. It was also possible to run the payload only with the TOF or tracker turned on.

\section{Flight}

\begin{figure}
\centerline{\includegraphics[width=1.0\linewidth]{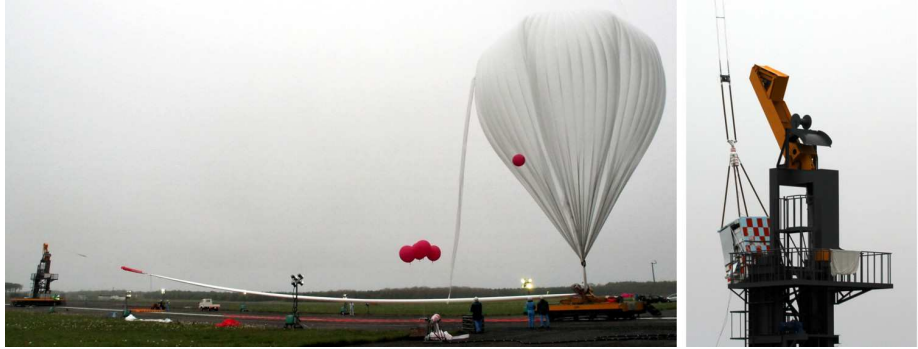}}
\caption{\label{f-fig4_jap_pgaps}\textbf{Left:} pGAPS on the launcher and inflated balloon right before launch. \textbf{Right:} pGAPS during take off.}
\end{figure}

\begin{figure}
\centerline{\includegraphics[width=1.0\linewidth]{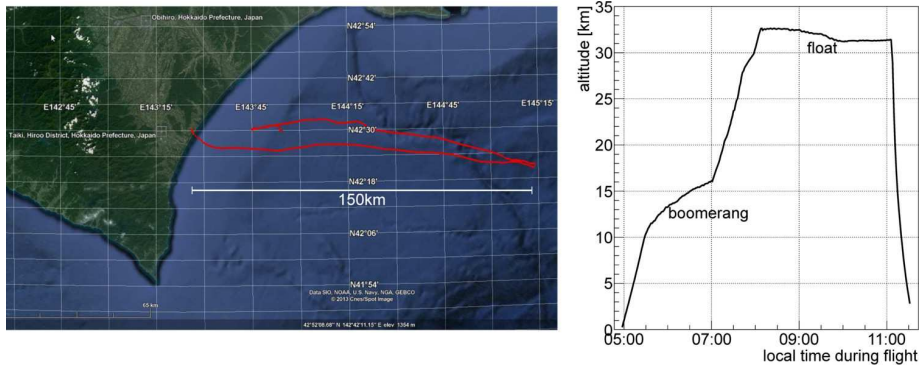}}
\caption{\label{f-fig5_jap_pgaps}\textbf{Left:} pGAPS balloon flight trajectory (Google Maps). \textbf{Right:} Altitude as a function of time during flight.}
\end{figure}

The scientific part of the pGAPS payload was assembled at the Space Sciences Laboratory of UC Berkeley starting in summer 2011 and was shipped to the Institute of Space and Astronautical Science/Japan Aerospace Exploration Agency (ISAS/JAXA) facilities in Sagamihara, Japan in April 2012 where it underwent thermal vacuum testing and was matched to the gondola bus. The bus gondola supplied power by Lithium batteries, telecommunication to ground, and ballast hoppers. The launch site was the Taiki Aerospace Research Field in Taiki at the southern tip of the east coast of Hokkaido, Japan and was preceded by compatibility testing, e.g., electromagnetic interference and communication, final rigging procedures, and a dress rehearsal \cite{taiki}.

The launch took place at 4:55\,am JST on June 3rd, 2012 using a FB-100 helium balloon with a volume of 100,000\,m$^3$ (Figure~\ref{f-fig4_jap_pgaps}). After the initial ascent the balloon drifted eastward for about three hours at 10--15\,km altitude (boomerang altitude) before dropping more ballast and further ascending to a maximum of \tilde33\,km (float altitude) \cite{jaxbal}. At that time of the season the winds at high altitude blew the balloon westward back to the coast of Hokkaido. The balloon was released at 11:05\,am and the payload landed in the water at 11:36\,am where it was recovered by boat within 9\,min (Figure~\ref{f-fig5_jap_pgaps}).

\subsection{Data taking and instrument health}

\begin{figure}
\centerline{\includegraphics[height=0.41\linewidth]{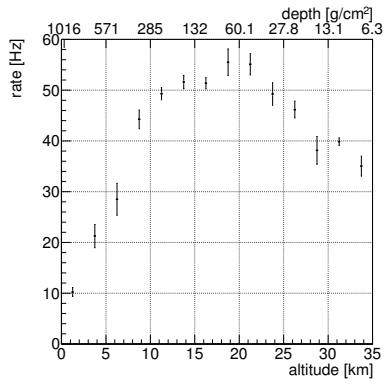}}
\caption{\label{f-fig6_jap_pgaps}TOF trigger rate during flight as a function of altitude and atmospheric depth.}
\end{figure}

Before launch the Si(Li) detectors were cooled down and calibration data with atmospheric muons and X-rays were recorded. pGAPS took science data, i.e., energy depositions in the tracker and energy depositions and time values from the TOF, from launch until the balloon was released, while housekeeping data were also recorded throughout descent. The 6:10\,h of science data taking were split into cadences of TOF trigger mode ($19\times13$\,min), X-ray tube calibration ($13\times4$\,min), tracker trigger mode ($9\times3$\,min), and at the very end TOF trigger mode with tracker turned off for 13\,min. These different modes made up 91\,\% of the flight and consist of data with nominal values for currents and voltages, temperatures as well as electronics status. The number of TOF triggers recorded during all well-defined data taking modes was \tilde$8\cdot10^5$, the number of TOF events also triggering the tracker was \tilde$6\cdot10^5$, and the number of Si(Li) detector triggers during X-ray tube calibration was \tilde$2.7\cdot10^6$. The TOF took data continuously throughout the flight even when the tracker was operated in tracker trigger mode. The profile of the raw TOF trigger rate as a function of altitude and atmospheric depth was generated by sampling the TOF trigger rate over 2\,min intervals and filling these values together with the mean altitude of this sample into a two dimensional histogram (Figure~\ref{f-fig6_jap_pgaps}). The error bars reflect the error on the mean trigger rate for each altitude bin. The error bar length is affected by the number of available samples for each altitude bin, which is defined by the altitude change velocity, and the spread of trigger rate samples at the specific altitude. A maximum at \tilde18\,km is seen and is in good agreement with other measurements and air shower simulations \cite{planeto,phd}.

All electronics (power distribution devices, tracker and TOF electronics) worked very reliably and the flight computer rebooted only once during ascent from boomerang altitude to float altitude due to the changing grounding environment.

\subsection{Performance of the tracker}

\subsubsection{Livetime}

The livetime of the tracker electronics for processing events during TOF trigger mode was on ground very close to 100\,\%, at boomerang altitude 99.8\,\%, and at float 99.5\,\%. Only during the high rate calibration with the X-ray tube did the livetime for the detector with the largest exposure drop to 95\,\%.

\subsubsection{Energy calibration\label{s-trkcali}}

\begin{figure}
\begin{center}
\begin{minipage}[t]{.4\linewidth}
\centerline{\includegraphics[width=1.0\linewidth]{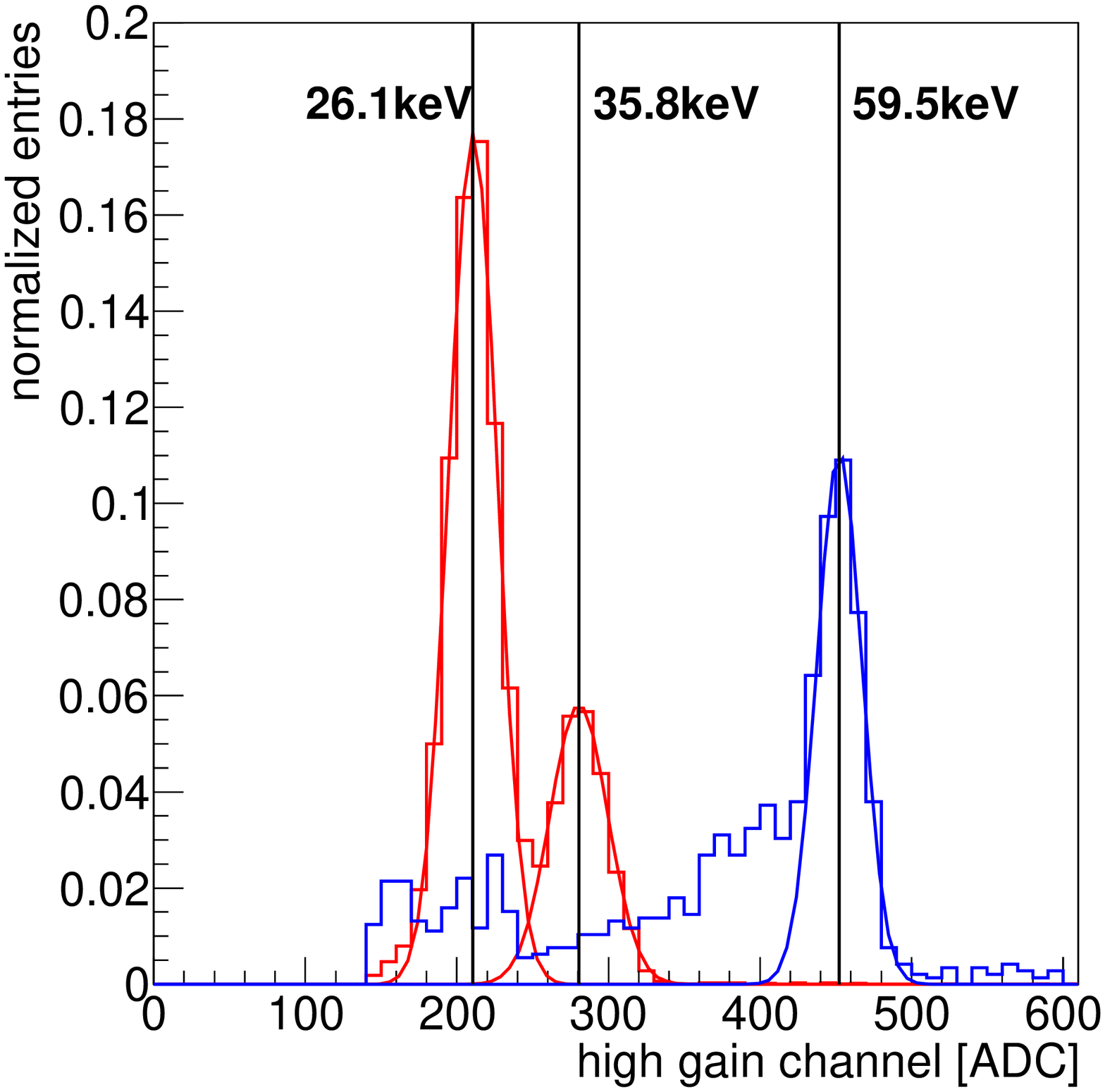}}\caption{\label{f-fig7_jap_pgaps}Calibration of one high gain channel with X-ray tube lines (red) and the 59.5\,keV Am-241 line (blue) for one of the detectors in the middle tracker layer (stack 0, layer 1, as defined in Figure~\ref{f-fig3_jap_pgaps}).}
\end{minipage}
\hspace{.1\linewidth}
\begin{minipage}[t]{.4\linewidth}
\centerline{\includegraphics[width=1.0\linewidth]{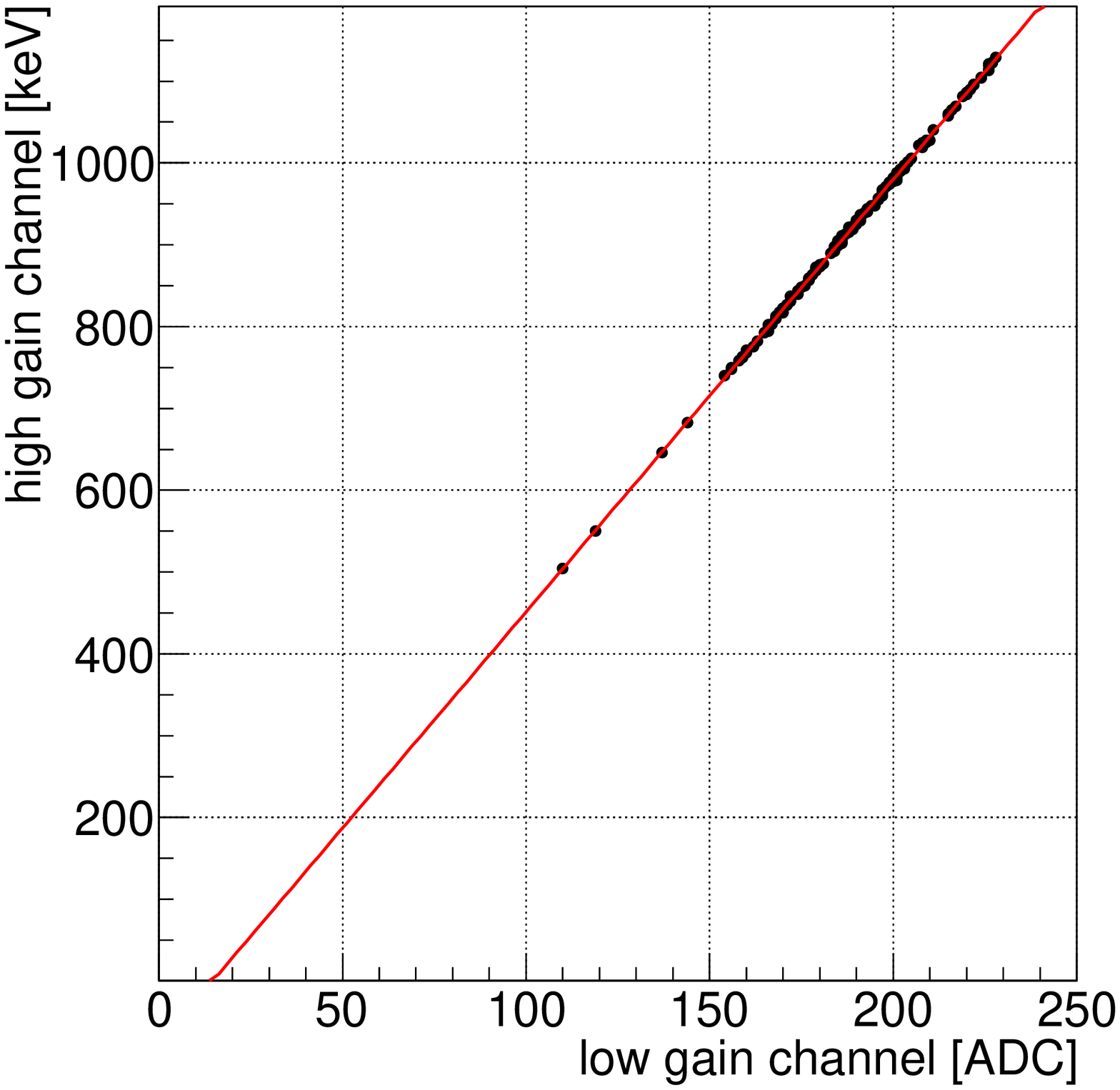}}\caption{\label{f-fig8_jap_pgaps}Calibration of low gain channel with the help of the overlap region between low and high gain channel.}
\end{minipage}
\end{center}
\end{figure}

The high gain readout channels were calibrated using the X-ray tube lines at 26 and 36\,keV and the Am-241 line at 59.5\,keV. Figure~\ref{f-fig7_jap_pgaps} shows the results for one typical channel during the preparation period a few hours before the launch. The histogram for the measurement with the Am-241 source shows a tail towards lower energies (45--55\,keV) that is assumed to be formed from scattered 59.5\,keV photons. The source could only be placed on top of the uppermost TOF layer before launch and scattering could occur in the detector material between the source and the detector. The three line positions with their corresponding widths were used as input to a straight line fit. The measurements with Am-241 could only be performed on ground as flying a radioactive source was not permitted. Laboratory measurements of the high gain readout channels with test pulses during the electronics development phase showed a linear behavior of the ADC response for the full 13\,bit range. Therefore, the low gain channel for the charged particle measurement was calibrated by studying the overlap region of the high and low gain channel connected to the same detector strip. Figure~\ref{f-fig8_jap_pgaps} shows one example for the correlation of the digitization of the same energy deposition between the already calibrated high gain channel and the analog-to-digital converter value (ADC) measured in the low gain branch. A clear linear correlation is visible and straight line fits for every channel were used to map the low gain ADC value to a calibrated energy value. 

\subsubsection{Stability and behavior as a function of temperature\label{s-trk_stab}}

\begin{figure}
\begin{center}
\begin{minipage}[t]{.4\linewidth}
\centerline{\includegraphics[width=1.0\linewidth]{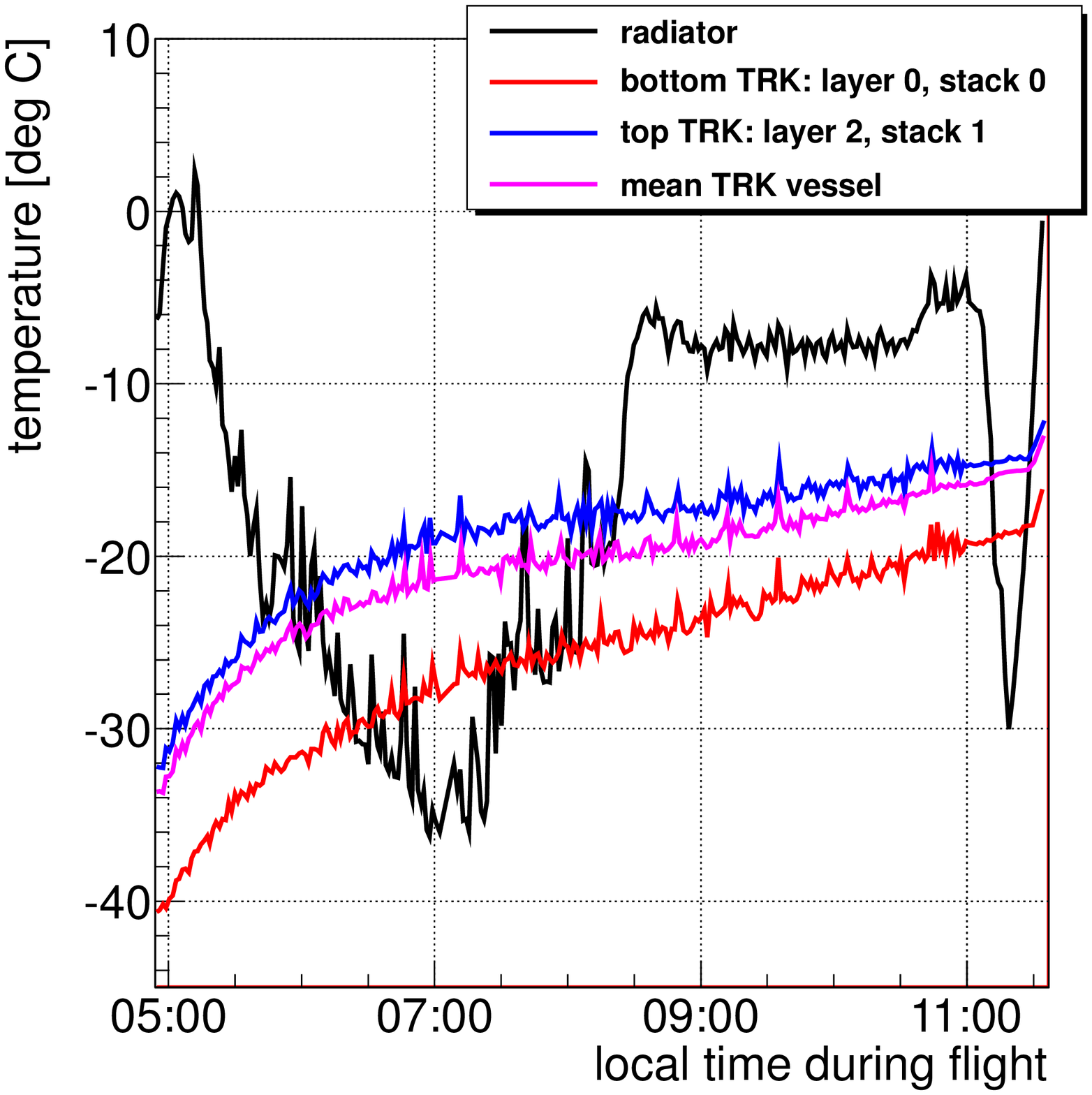}}\caption{\label{f-fig9_jap_pgaps}Temperatures as a function of time during flight (temperature of the radiator surface: black, temperature of the Si(Li) detector frame for one of the bottom/top Si(Li) modules: red/blue, mean temperature inside the Si(Li) detector vessel: magenta). }
\end{minipage}
\hspace{.1\linewidth}
\begin{minipage}[t]{.4\linewidth}
\centerline{\includegraphics[width=1.0\linewidth]{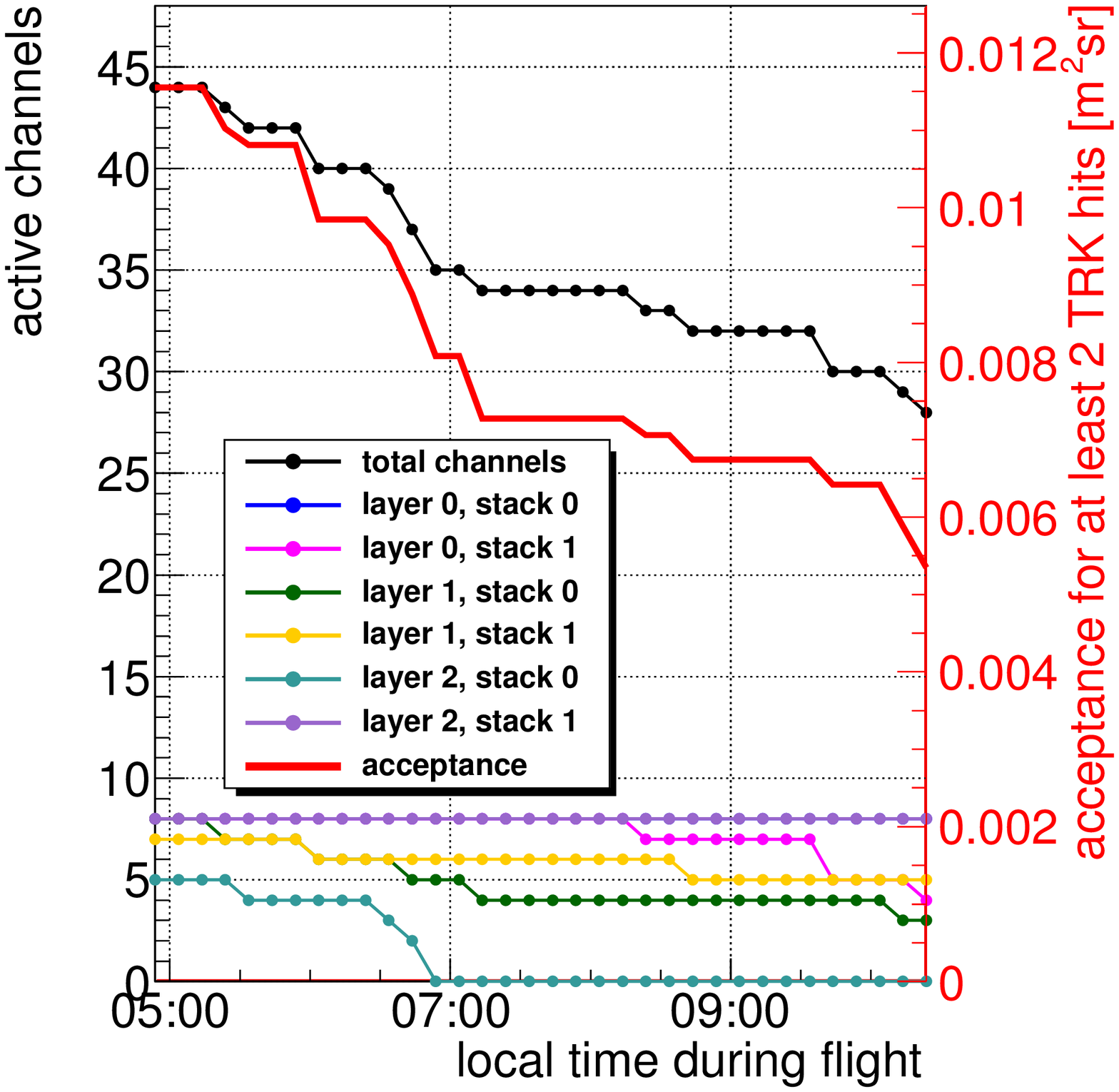}}\caption{\label{f-fig10_jap_pgaps}Active channels as a function of time during flight for individual Si(Li) detectors (blue, magenta, green, yellow, light blue, purple) and integrated over all detectors (black). In addition, the geometrical acceptance for TOF trigger events with at least two tracker hits in active channels is shown (red).}
\end{minipage}
\end{center}
\end{figure}

As mentioned above, a critical parameter for the operation of Si(Li) detectors is the temperature. Ground testing showed that most detector channels started to deplete from about -15\de~to -20\de. The coldest temperature at launch inside the tracker vessel measured on one of the detectors of the bottom layer was about -40\de~and the average temperature was -34\de. Figure~\ref{f-fig9_jap_pgaps} shows the temperature evolution of some of the relevant Si(Li) detector components during the flight. The radiator was directly exposed to the ambient medium. During ascent it was not planned to have attitude control as the atmosphere is too dense and therefore the gondola was spinning randomly. Upgoing spikes in the temperature distribution of the radiator are correlated with facing the Sun side. After reaching float altitude the attitude control system was supposed to point the radiator towards the anti-Sun side of the gondola to act as a heat dump for the tracker modules. Unfortunately, due to an operational mistake this control was not possible. As a result the gondola was spinning at an approximately constant rate of one rotation every 5\,min, as measured by the fiber optical gyroscope, causing the tracker vessel to warm up by \tilde0.02\de/min. One of the main goals of the pGAPS flight was to verify the thermal model for the cooling approach of the Si(Li) detectors. Although the rotator failed, this goal was fully accomplished and the thermal analysis of all recorded temperature sensors led to a complete understanding of the thermal system \cite{isaacpaper}. For flights of the full GAPS instrument from Antarctica the well established rotator from the Columbia Scientific Balloon Facility would be used.

However, the effect of the increasing tracker temperature needs to be accounted for in the science flight data analysis. Figure~\ref{f-fig10_jap_pgaps} shows the time evolution of the total number of active channels for the individual detectors over the course of the flight. Under nominal operating conditions, out of 48 total channels 44 channels were operating while the other four channels were known not to be working since the qualification phase. At the time the tracker was turned off (\tilde10:30\,am) the mean temperature was about -17\de~and 28 channels were still operational while the other 16 channels were either non-functional due to high leakage currents saturating the preamplifier or being no longer depleted. Laboratory measurement after the flight at nominal temperatures showed that these channels were fully functional. The individual detectors and also the strips within a detector showed a variation with temperature, which is explainable by two different effects. Some of the wire bonds between the Si(Li) detector printed circuit boards to the detector surfaces were of visibly poor quality, which increased the noise level and therefore required lower operating temperatures compared to neighboring strips with better bonding quality. Another effect was related to the surface condition of the grooves between the strips causing channels to deplete at different voltages due to inter-electrode capacitance differences. These effects were carefully investigated as for the full-scale payload the Si(Li) modules will be fabricated in house by the GAPS collaboration. For instance, one advantage of the GAPS Si(Li) modules over the Semikon modules is the use of a robust pressure contact instead of wire bonding. Another advantage is that the detectors will be structured on the n+ side instead of the p+ side. In this way, strips can be more easily separated as the depletion starts from the n+ side.

\begin{figure}
\centerline{\includegraphics[width=1.0\linewidth]{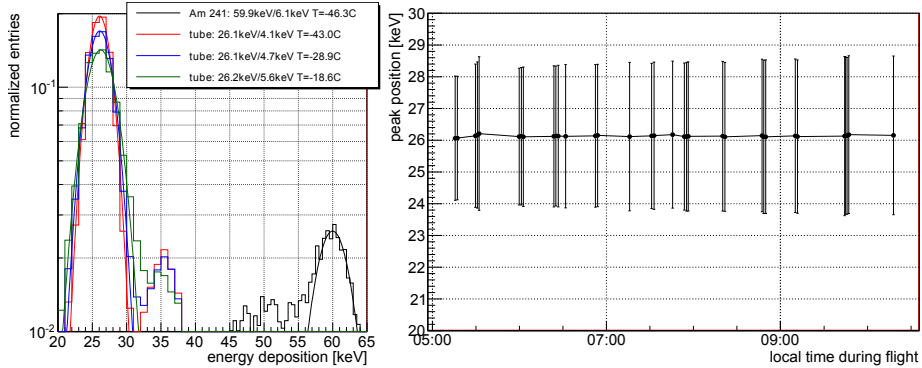}}
\caption{\label{f-fig11_jap_pgaps}\textbf{Left:} X-ray tube (preparation/red, right after launch/blue, float/green) and Am-241 source (preparation/black) measurements for one detector (stack 0, layer 0, as defined in Figure~\ref{f-fig3_jap_pgaps}) integrated over all eight channels. Energy values printed in the legend denote the mean and the FWHM of the Gaussian fit. \textbf{Right:} Time evolution of the peak position of the Gaussian fit for the 26\,keV X-ray tube peak. The error bars denote the widths of the Gaussians.}
\end{figure}

The left side of Figure~\ref{f-fig11_jap_pgaps} shows the spectra for the X-ray energy deposition calibration for the detector closest to the X-ray tube. The spectra are integrated over all eight channels. The X-ray tube spectra were recorded during the launch preparation, right after launch, and after reaching float altitude. In comparison, and as already introduced in Section~\ref{s-trkcali}, the Am-241 spectrum taken a few hours before launch is shown. Compared to the measurement shown for a different detector in Figure~\ref{f-fig7_jap_pgaps}, a weaker Am-241 line with respect to the X-ray tube lines is visible because less material is in front of the X-ray tube and more material in front of the Am-241 source. The right side of Figure~\ref{f-fig11_jap_pgaps} depicts the evolution of the position and width of the Gaussian fit for the dominant 26\,keV X-ray tube peak over time. A small broadening is visible, which can be explained by a temperature increase of 11.5\de~over the course of the flight, which translates into an increase of the full width half maximum value (FWHM) of \tilde0.08\,keV/\de. A similar measurement was carried out during the integration period using the Am-241 source where an average of ($0.08\pm0.05$)\,keV/\de~in the temperature range from -40\de~to -20\de~was observed. In conclusion, the observed effect of energy resolution change with temperature is small. Once a strip was cold enough (\tilde-20\de) to be depleted and have a small enough leakage current the X-ray resolution improved only by about 10-15\,\% until it reached the nominal operational temperature of -35\de. Therefore, the temperature dependence would not significantly affect the ability to separate X-rays from antiprotonic exotic atoms from antideuteronic exotic atoms. The energy depositions of charged particles are much higher (\tilde1\,MeV) and behaved very stably.

However, a good temperature regulation is critical to keep all detector channels at low leakage currents so as to not have holes in the acceptance. The effect of the increasing number of non-operational channels on the geometrical acceptance for pGAPS for charged particle tracks with at least two tracker hits in TOF trigger mode over the course of the flight is shown in Figure~\ref{f-fig10_jap_pgaps} and will be further discussed in Section~\ref{s-accep}.

\subsubsection{Si(Li) track analysis}

\begin{figure}
\begin{center}
\begin{minipage}[t]{.4\linewidth}
\centerline{\includegraphics[width=1.0\linewidth]{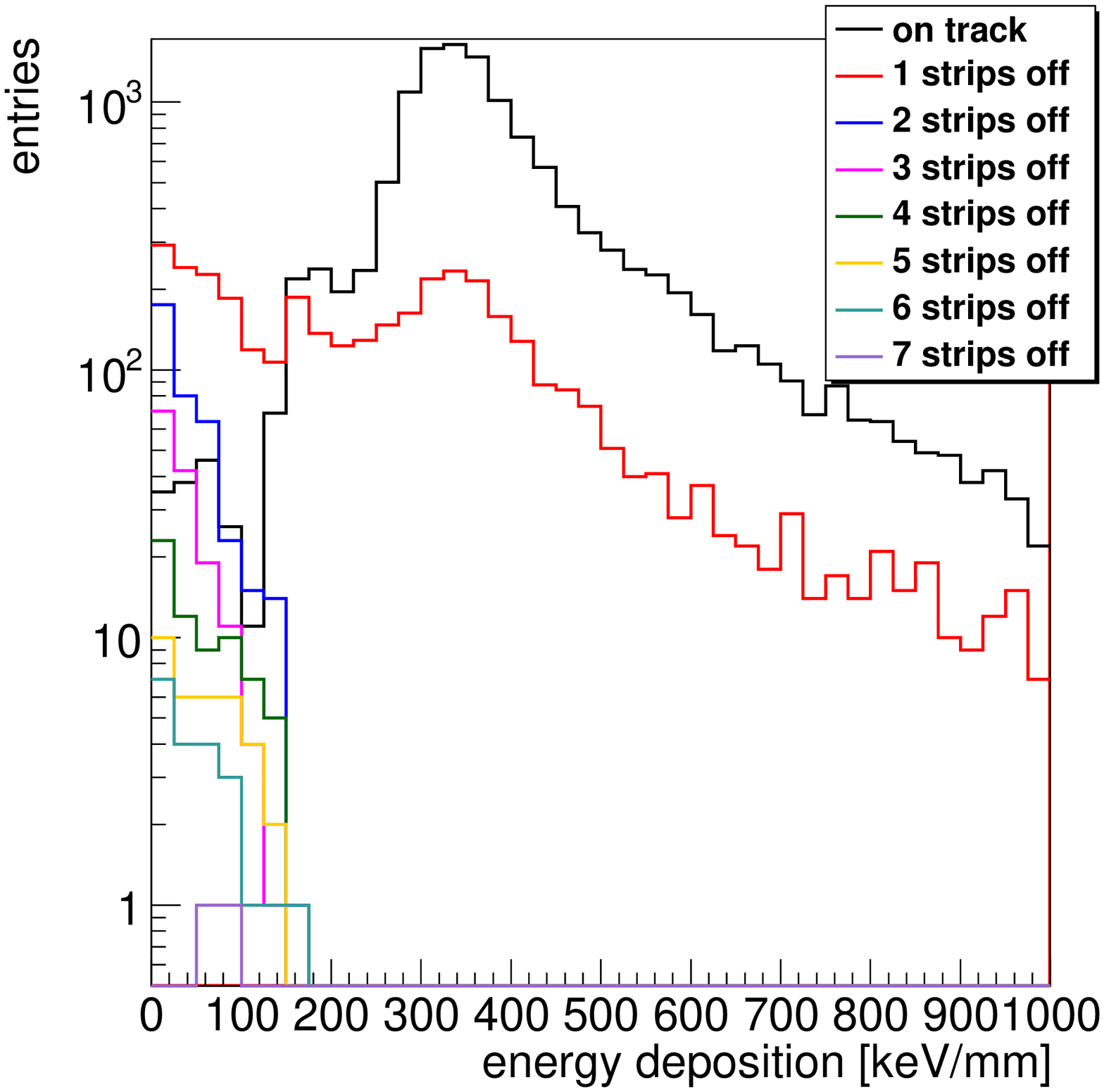}}\caption{\label{f-fig12_jap_pgaps}Energy depositions per mm in the Si(Li) material for different distances to the track fit position in number of strips integrated over all detectors.}
\end{minipage}
\hspace{.1\linewidth}
\begin{minipage}[t]{.4\linewidth}
\centerline{\includegraphics[width=1.0\linewidth]{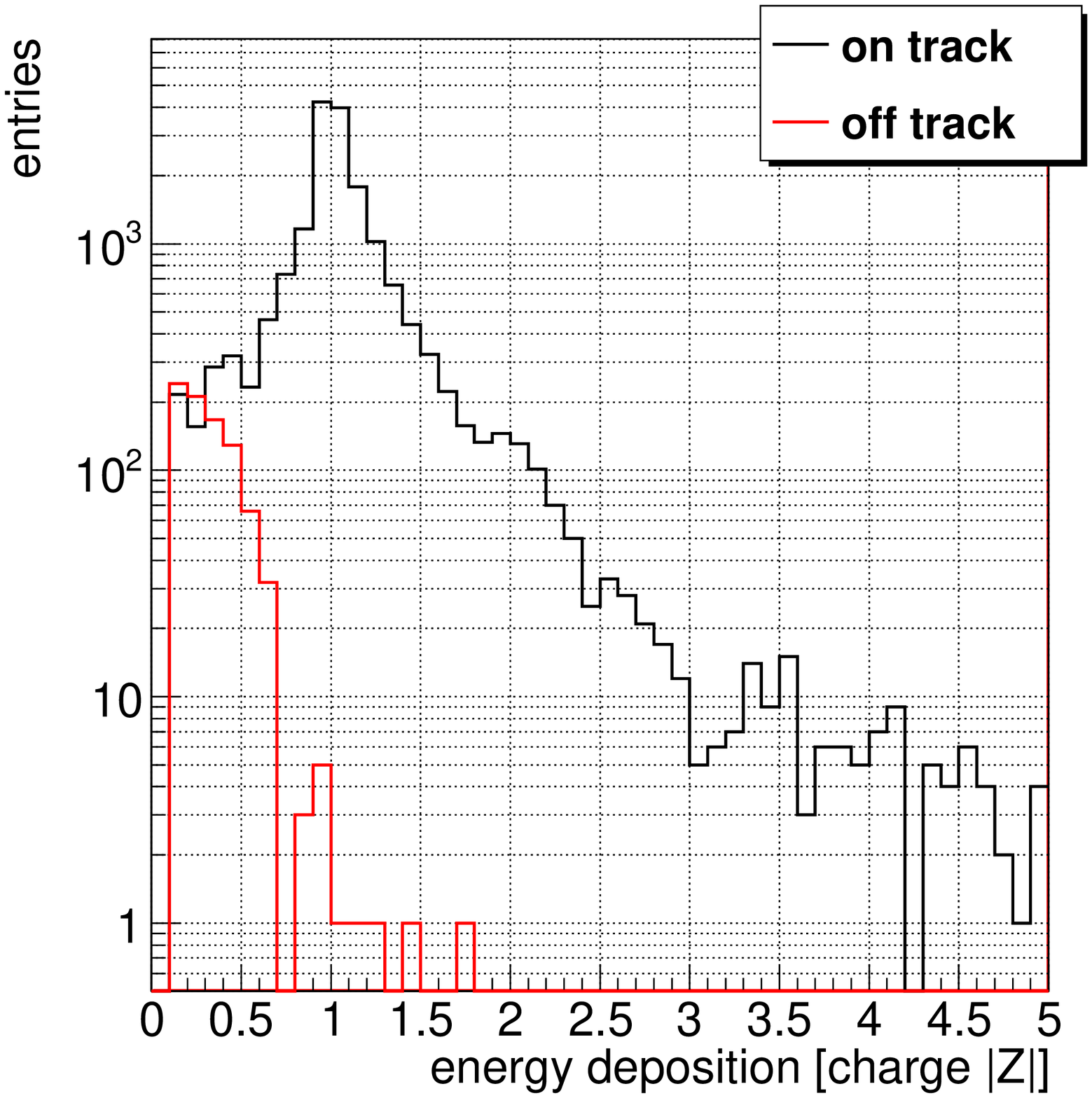}}\caption{\label{f-fig13_jap_pgaps}Energy depositions normalized to charge $|Z|$ for on (black) and off (red) track hits.}
\end{minipage}
\end{center}
\end{figure}

As outlined before, the TOF trigger was set up in such a way to either trigger on a combination of top and middle layers or on a combination of middle and bottom layers. For pGAPS about 10\,\% of the charged particle triggers included at least one hit in the Si(Li) modules. A track fit incorporating both TOF and tracker hits was performed and is explained in detail in \ref{s-track}. Figure~\ref{f-fig12_jap_pgaps} illustrates the energy deposition $E\sub{dep}$ distributions in the Si(Li) detectors for different distances to the actual hit position for all reconstructed events during the flight integrated over all detectors. The distributions are normalized to the path length in the material:\be\frac{\text d E\sub{dep}}{\text d x}=\frac{E\sub{dep}}{d}\cos\theta,\ee where $\theta$ is the zenith angle of the track and $d$ the thickness of the individual Si(Li) module. The black distribution shows the behavior for strips on the track with a clear peak at ($341\pm10$)\,keV/mm, a steep decline towards smaller energy depositions, and a long tail, as expected from a Landau distribution for charged particle energy depositions. The red distribution represents the energy depositions for adjacent strips to the strip position calculated from the track parameters. It is very similar in shape to the on-track distribution starting from the most probable value (MoP). The resolution of the track fit is about 1.3 Si(Li) strips wide and therefore allows hits in the neighboring strip of the calculated position. A smaller fraction of this distribution can also be attributed to cross-talk between adjacent strips due to capacitive coupling (Section~\ref{s-trk_stab}). Below the MoP more low energy entries are visible than for the on-track distribution. Contributing effects are increased detector noise from the non-optimal tracker operating temperature (Section~\ref{s-trk_stab}), coherent atmospheric shower particles and lower energy depositions from interaction products of the primary charged particle in the detector material moving at a different angle, or to a very small degree incoherent atmospheric and cosmic X-rays. The distribution shapes for strips further than one strip away from the calculated track is radically different and only show a low energy contribution, supporting the reliability of the track fit. In the following, strips adjacent to the calculated track position with non-zero energy depositions are considered as on-track hits and farther away strips as off-track. 

As the MoP for a minimum ionizing charged particle energy distribution scales as the square of its charge $Z$ and the majority of all charged particles during the flight were expected to have an absolute charge value of $|Z|=1$, the $\text dE\sub{dep}/\text d x$ values are transformed to absolute charge values $|Z|$ by:\be|Z|=\sqrt{\frac{\text dE\sub{dep}/\text d x}{341\,\text{keV/mm}}} \ee and the corresponding distribution for all on and off track hits in the Si(Li) during the flight is shown in Figure~\ref{f-fig13_jap_pgaps}. The on-track distribution demonstrates a clear peak at $|Z|=1$ and an additional shoulder at $|Z|=2$ due to $\alpha$ particles. The particle composition will be further discussed in Section~\ref{s-charge}.

In addition, the track fit information allows the extraction of the detection efficiency. Using only the cleanest subsample of tracks with no off-track hits in TOF or Si(Li), requiring at least two Si(Li) hits in different layers, tracks going through the inner part of a module up to a radius of 3.7\,cm, and neglecting non-operational Si(Li) strips, the tracker showed an average detection efficiency of (95.3$\pm$0.2)\,\%. This value should be interpreted as a lower bound as the Si(Li) module operating temperature was not optimal.

\subsubsection{X-ray backgrounds}

\begin{figure}
\begin{center}
\begin{minipage}[t]{.4\linewidth}
\centerline{\includegraphics[width=1.0\linewidth]{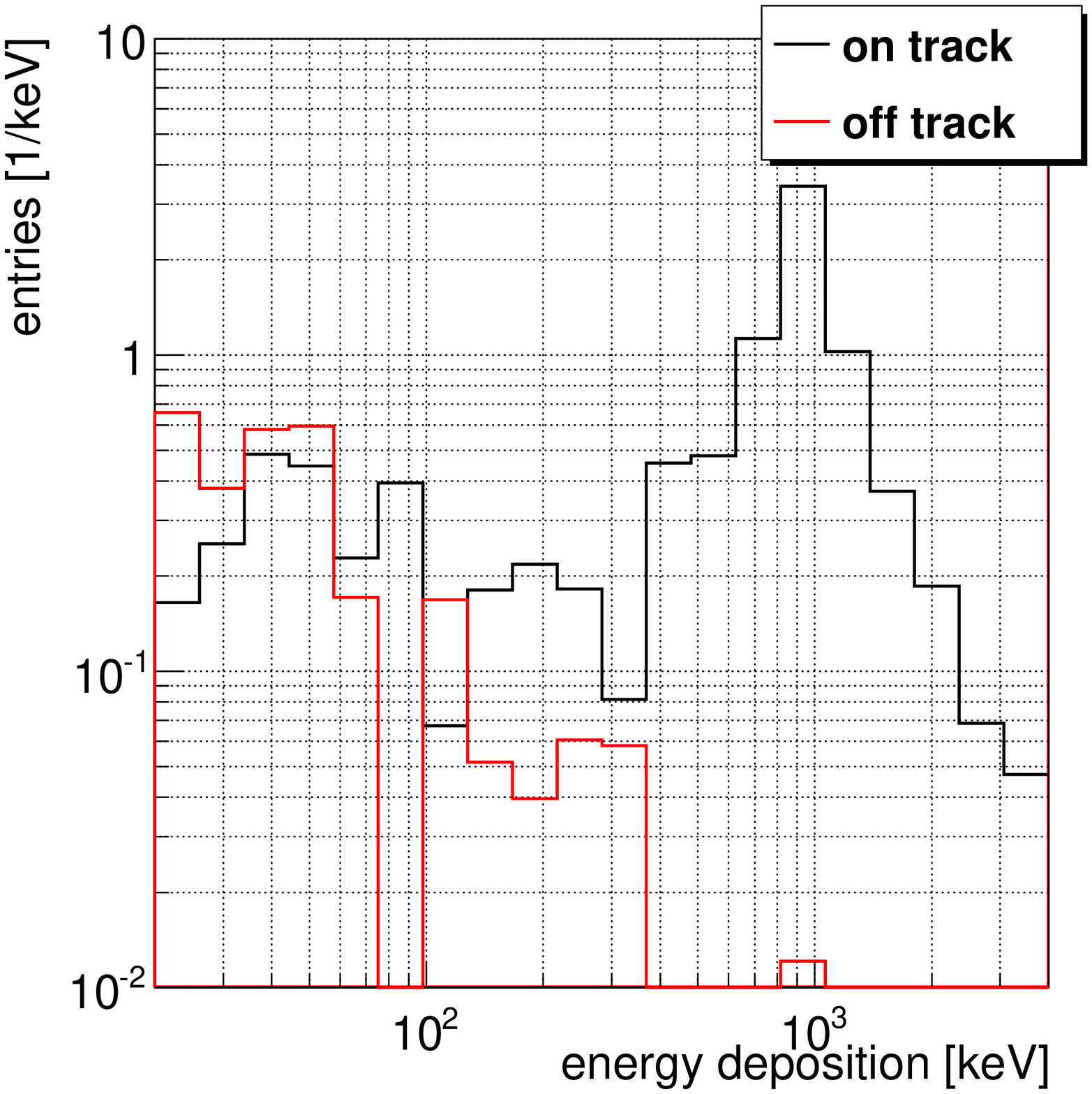}}\caption{\label{f-fig14_jap_pgaps}Energy depositions in keV for strips on track (black) and off track (red) during TOF trigger mode.}
\end{minipage}
\hspace{.1\linewidth}
\begin{minipage}[t]{.4\linewidth}
\centerline{\includegraphics[width=1.0\linewidth]{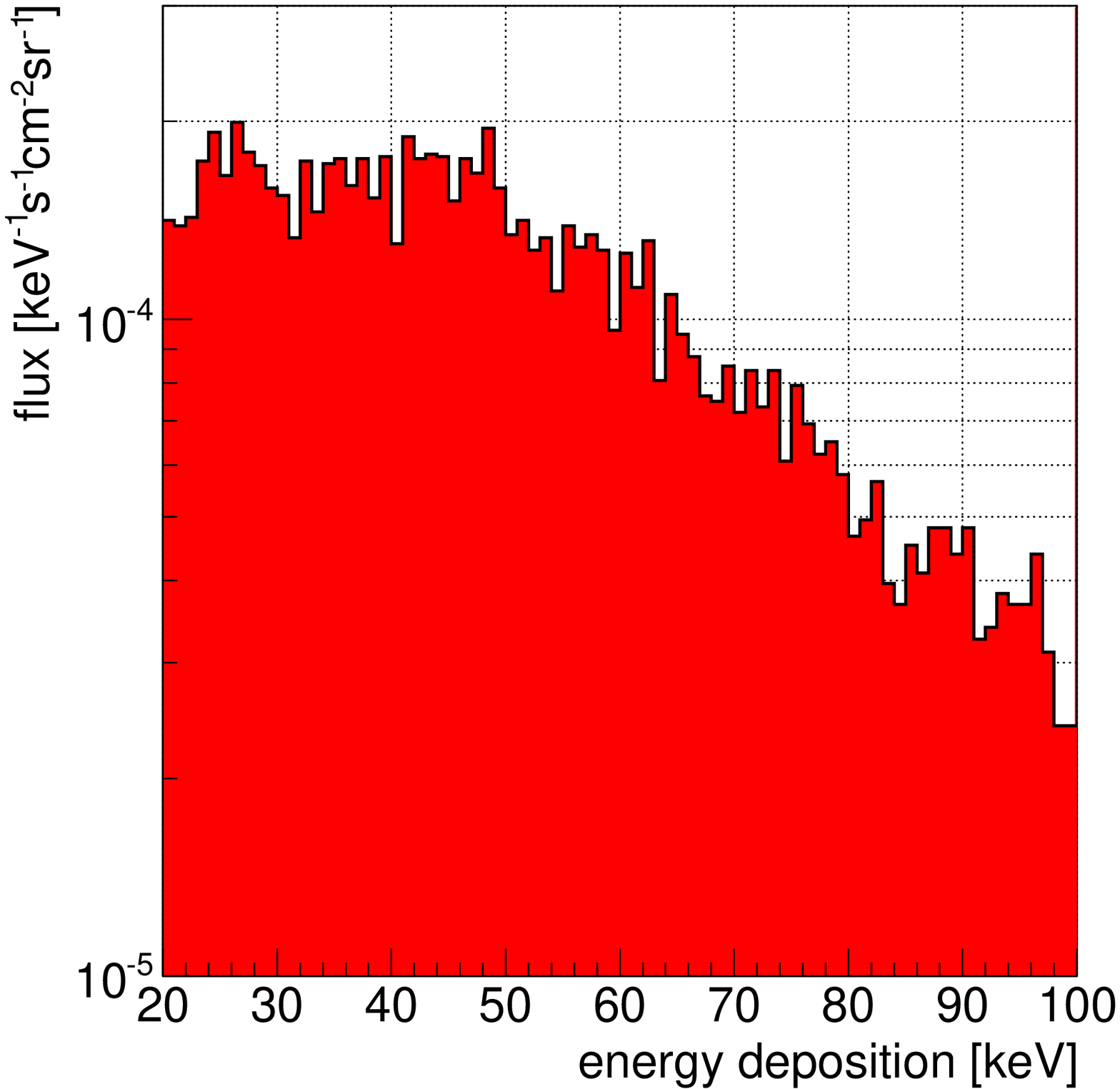}}\caption{\label{f-fig15_jap_pgaps}X-ray flux during tracker trigger mode.}
\end{minipage}
\end{center}
\end{figure}

As the GAPS antideuteron search will rely on a good X-ray identification, it is important to understand the level of coherent and incoherent X-ray backgrounds in coincidence with charged particles, e.g., from particle interactions or atmospherically produced X-rays. For this study only the detector with the weakest temperature dependence was used. The critical energy range for the antideuteron analysis is between 20 and 100\,keV and both the on and off track distributions show a very similar behavior in this range (Figure~\ref{f-fig14_jap_pgaps}). The statistics for this study in TOF trigger mode are low, but the data from the tracker trigger mode without the X-ray calibration tube can be used to better understand the atmospheric and cosmic X-ray component. In this regard, Figure~\ref{f-fig15_jap_pgaps} illustrates the X-ray energy deposition flux at an average float altitude of about 32\,km using a geometrical detector acceptance of 436\,cm$^2$sr (both sides) and 27\,min of livetime. The coincidence flux of X-rays and charged particles will be presented after the charged particle flux discussion in Section~\ref{s-chargex}.

\subsection{Performance of the time-of-flight detector}

\subsubsection{Livetime\label{s-tof-live}}

\begin{figure}
\centerline{\includegraphics[width=0.4\linewidth]{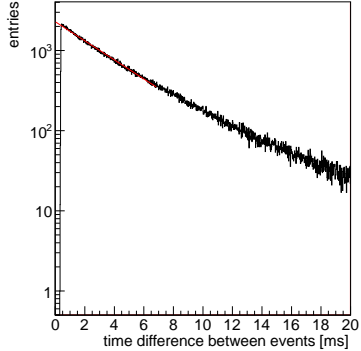}}
\caption{\label{f-fig16_jap_pgaps}Time difference between TOF triggers of successive events during TOF trigger mode.}
\end{figure} 

The livetime of the TOF system is defined by the electronics deadtime of 280\,\textmu s. Figure~\ref{f-fig16_jap_pgaps}  shows the distribution of time differences between successive raw TOF trigger events. As expected an exponential behavior is observed with a gap of 280\,\textmu s length in the beginning. The livetime is estimated by fitting an exponential to the distribution and calculating how many events were missed during the electronics processing period and results in $(93.2\pm0.1)$\,\%.

\subsubsection{Energy deposition calibration and stability}

\begin{figure}
\centerline{\includegraphics[width=1.0\linewidth]{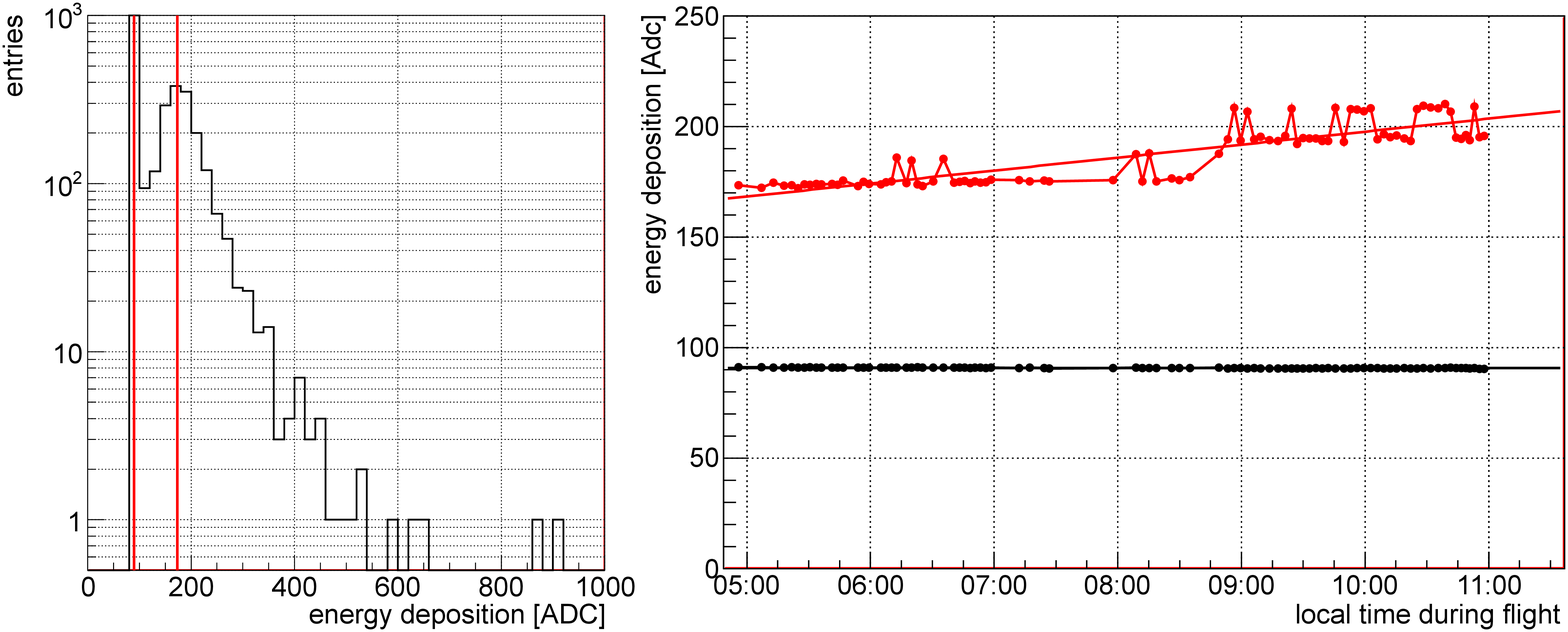}}\caption{\label{f-fig17_jap_pgaps}\textbf{Left:} Pedestal and most probable value positions for one typical TOF PMT. The vertical red lines indicate the pedestal and MoP value position, respectively. \textbf{Right:} Behavior of the most probable (red) and pedestal (black) value over the course of the flight for the same PMT.}
\end{figure}

The energy deposition in each paddle was measured by two PMTs and digitized with charge sensitive ADCs. The raw ADC spectrum of one PMT shows a constant offset (pedestal) followed by a Landau distribution for the actual energy deposition (Figure~\ref{f-fig17_jap_pgaps} left). For each PMT, a search to determine the pedestal and most probable value of the Landau distribution was performed. As discussed above, the MoP position scales as $Z^2$ for $Z$ being the charge of the incident particle. The search was sampled over the flight to study the stability of the pedestal and MoP positions (Figure~\ref{f-fig17_jap_pgaps} right). While the pedestal value $ADC\sub{ped}$ was very stable, the MoP value $ADC\sub{MoP}$ showed a slight increase over the course of the flight and was fitted with a straight line. Each measured PMT ADC energy deposition was calibrated to $|Z|$ by performing:

\be
|Z|=\sqrt{\frac{\displaystyle ADC-ADC\sub{ped}}{\displaystyle ADC\sub{MoP}(T)-ADC\sub{ped}}}.
\ee

\begin{figure}
\centerline{\includegraphics[width=0.4\linewidth]{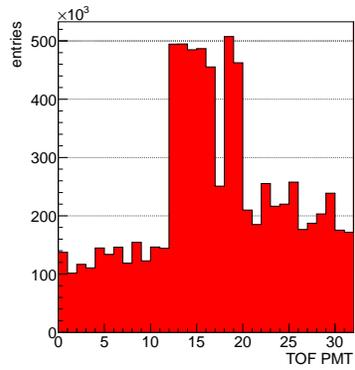}}
\caption{\label{f-fig18_jap_pgaps}Occupancy of TOF PMTs. Bottom layer: PMT 0-11, middle layer: PMT 12-19, top layer: PMT 20-31.}
\end{figure} 

Similar to the previous tracker discussion, the following analysis will also make use of the path length corrected energy deposition: $|Z|\sub{path}=|Z|/d\cos\theta$. Despite the necessity of a slightly time dependent ADC calibration, the TOF operation was stable over the course of the flight for 31 out of 32 PMTs. Figure~\ref{f-fig18_jap_pgaps} depicts the PMT occupancy for the whole flight for energy depositions clearly above the pedestal ($|Z|>0.3$). As it was a trigger requirement that the middle layer was always part of an event, the middle PMTs show about twice as many entries as the top and bottom layer PMTs. The distance between the top and middle layer was 18\,cm shorter than the distance between middle and bottom, and therefore the top PMTs show more entries than the bottom PMTs. All TOF PMTs operated for the full duration of the pGAPS flight, except for one tube in the middle TOF layer. This PMT showed normal operation on the ground and in the flight while the ambient pressure stayed above \tilde40\,torr. As soon as the pressure dropped below this value, the tube HV became unstable. This is consistent with corona discharge from exposure of part of the HV circuit to the low pressure ambient environment. Since the behavior manifested as soon as the pressure was low enough (and not after a period of outgassing), it suggests that the base assembly and potting was compromised some time after PMT vacuum testing. The PMT assemblies were integrated into the instrument about one year before the flight, and the paddle that this particular tube was mounted on was removed (along with the other three middle layer paddles) and re-installed every time the Si(Li) detector vessel was accessed. Therefore, it was subjected to some handling stress on a somewhat frequent basis.

Obviously in an experiment such as a GAPS science flight, a more rigorous testing regime will be needed. GAPS will have hundreds instead of dozens of PMTs, and the flight time will be months. PMT assemblies for GAPS will be tested in a low pressure environment for at least a week prior to acceptance. Thermal-vacuum testing for at least a subset of assemblies will be considered as well. Finally, some strengthening and enhanced strain relief of the PMT assembly is possible, especially for the cable feedthroughs.

\subsubsection{Timing calibration}

\begin{figure}
\begin{center}
\begin{minipage}[t]{.4\linewidth}
\centerline{\includegraphics[width=1.0\linewidth]{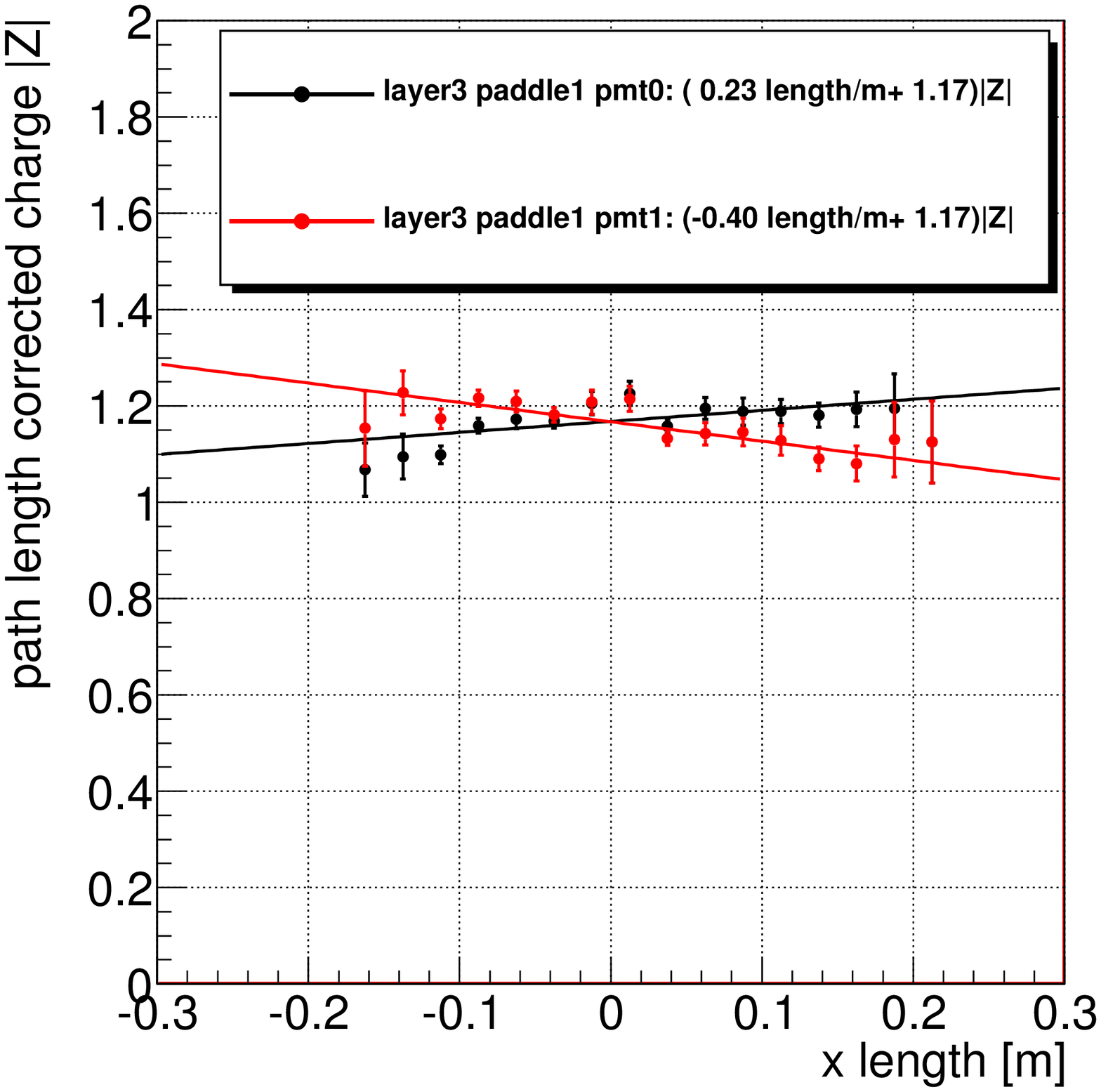}}\caption{\label{f-fig19_jap_pgaps}Path length corrected energy deposition as a function of hit position along a paddle. The black and red data points/straight line fits show the behavior of both PMTs attached to the same paddle.}
\end{minipage}
\hspace{.1\linewidth}
\begin{minipage}[t]{.4\linewidth}
\centerline{\includegraphics[width=1.0\linewidth]{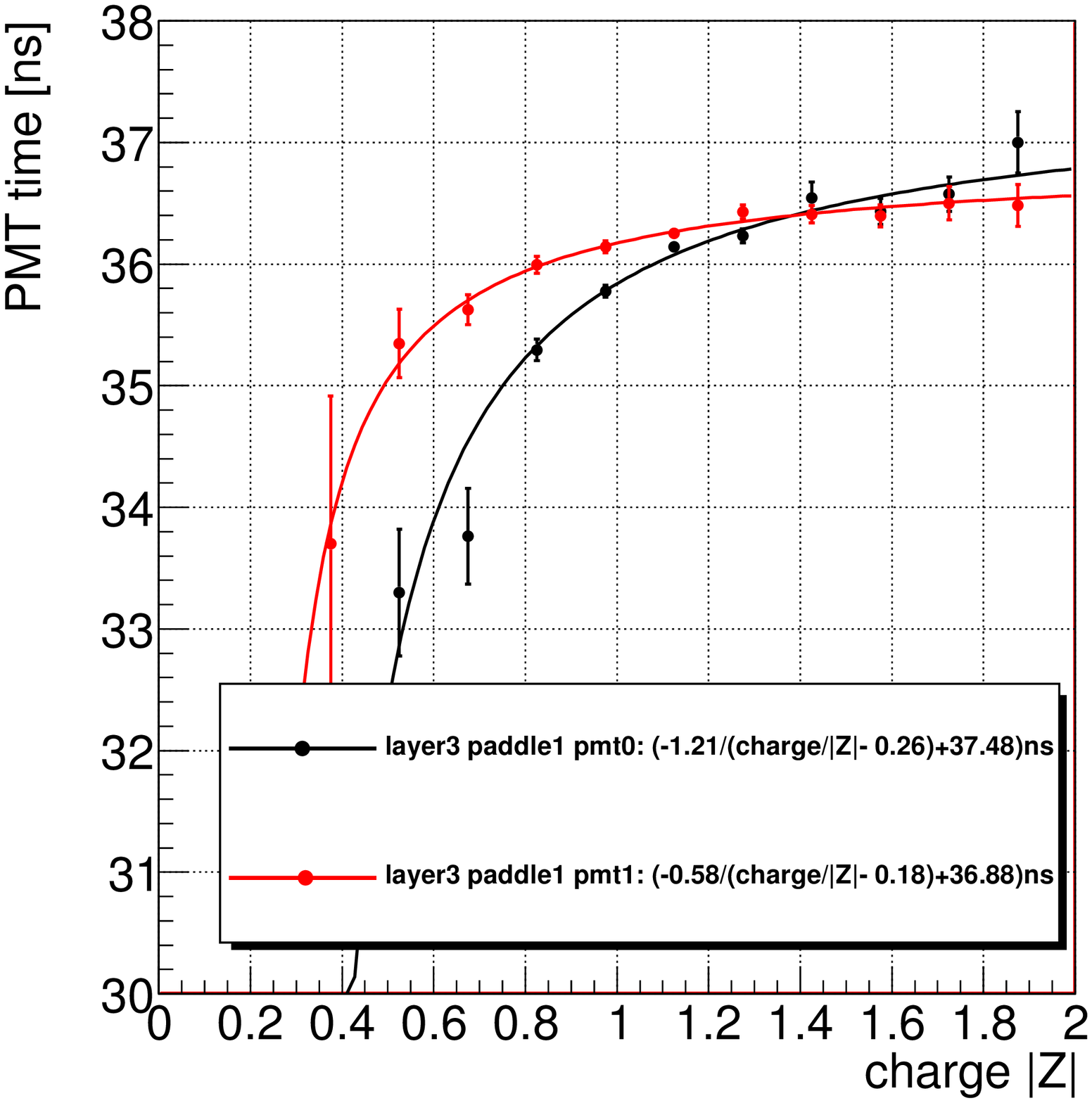}}\caption{\label{f-fig20_jap_pgaps}TDC time information as a function of energy deposition for the center slice of the paddle ($10\times15$\,cm). The black and red data points/hyperbolic fits show the behavior of both PMTs attached to the same paddle.}
\end{minipage}
\end{center}
\end{figure}

In addition to providing the main trigger for pGAPS, the time-of-flight was also designed to measure charged particle velocities. A TDC value was recorded for a PMT when the energy deposition pulse crossed a preset discriminator threshold. The digitization had a precision of one count per 50\,ps. The velocity measurement requires an inter-calibration of all TDCs to reliably measure time of flights. Important effects that need to be accounted for are the pulse propagation from the hit position to the PMT and the dependence of the TDC hit measurement on the pulse size at the PMT. All figures shown in this section depict the PMTs from the same paddle. The steps described in the following were carried out for each paddle. To deconvolve the different effects the pGAPS tracking information is used. Figure~\ref{f-fig19_jap_pgaps} shows for both PMTs the path length corrected energy deposition calibrated to charge $|Z|\sub{path}$ as a function of the hit position calculated from the track parametrization for the coordinate along the paddle. The data points were fitted with straight lines and the signal is about 5-10\,\% larger directly in front of the PMT compared to the center of the paddle. As expected, the lines for PMTs on opposite sides cross in the center. As a first step in the timing calibration, all mean $|\bar Z|\sub{path}$ values in the center of each paddle were scaled to the same value (mean of all uncalibrated paddle $|Z|\sub{path}$ values in the center). Using the same scaling factor $f$ but not the path length correction, the measured time as a function of $|Z|$ for the $10\times15$\,cm slice in the center of a paddle for both PMTs is shown in Figure~\ref{f-fig20_jap_pgaps}. The data points are well fitted with a hyperbolic function and reveal that the TDC value dependence on the energy deposition is rather mild for signals above \tilde0.8$|Z|$ and nearly constant starting from about 1.2$|Z|$. For smaller signals the slope becomes much steeper. To decouple the measured TDC value $t$ from the pulse height the TDC hits are corrected to the same $|Z|$ value using the hyperbolic fit:
\be t\sub{corr}=t-a_i\left(\frac{1}{|\bar Z|-b_i}-\frac{1}{|Z|\cdot f_i-b_i}\right)\ee with $a_i$, $b_i$ being the fit values from the hyperbolic fit and $f_i$ the energy deposition scaling factor for each PMT $i$.

\begin{figure}
\begin{center}
\begin{minipage}[t]{.4\linewidth}
\centerline{\includegraphics[width=1.0\linewidth]{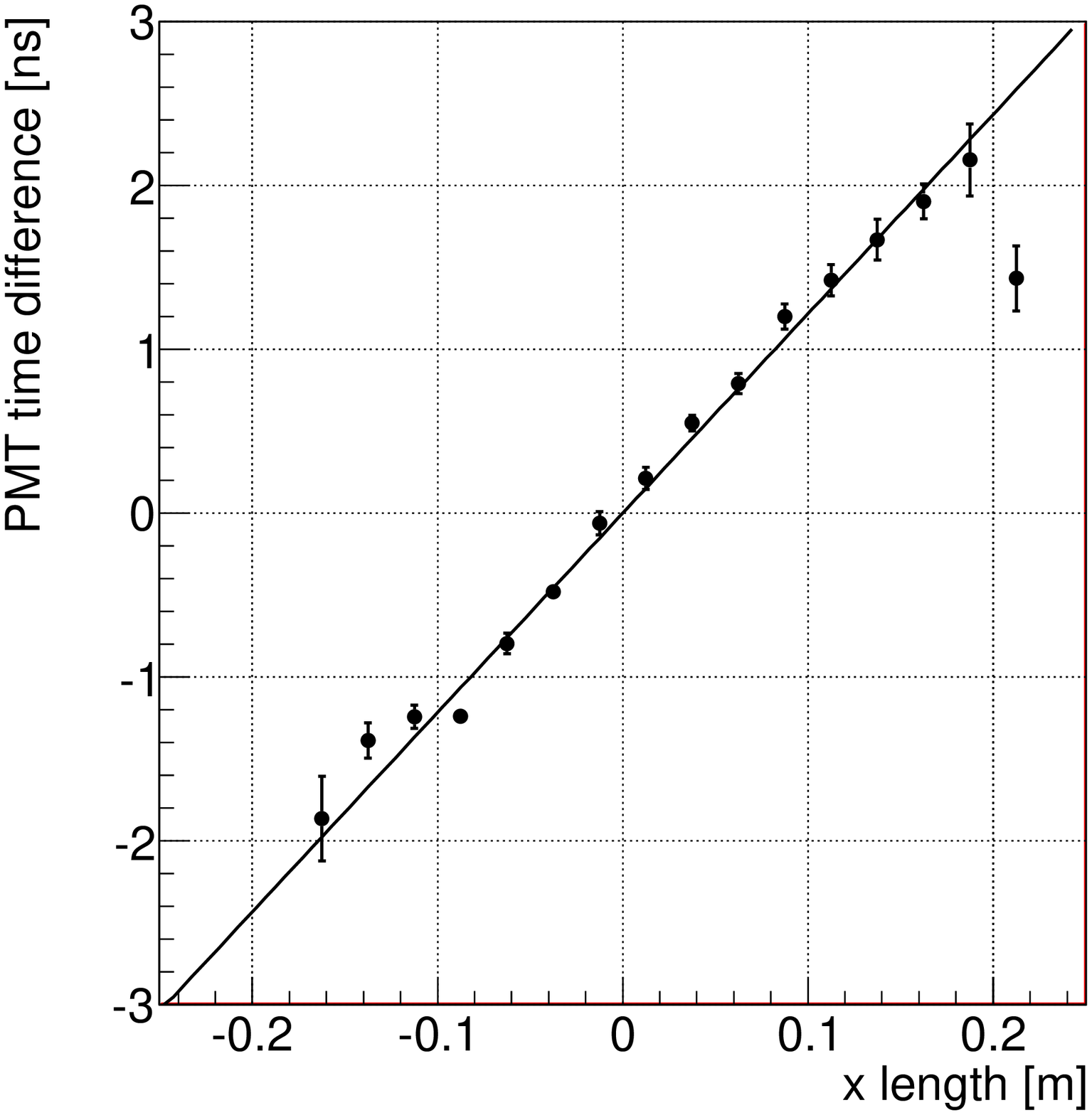}}\caption{\label{f-fig21_jap_pgaps}TDC time difference of both PMTs attached to a paddle as a function of hit position along a paddle.}
\end{minipage}
\hspace{.1\linewidth}
\begin{minipage}[t]{.4\linewidth}
\centerline{\includegraphics[width=1.0\linewidth]{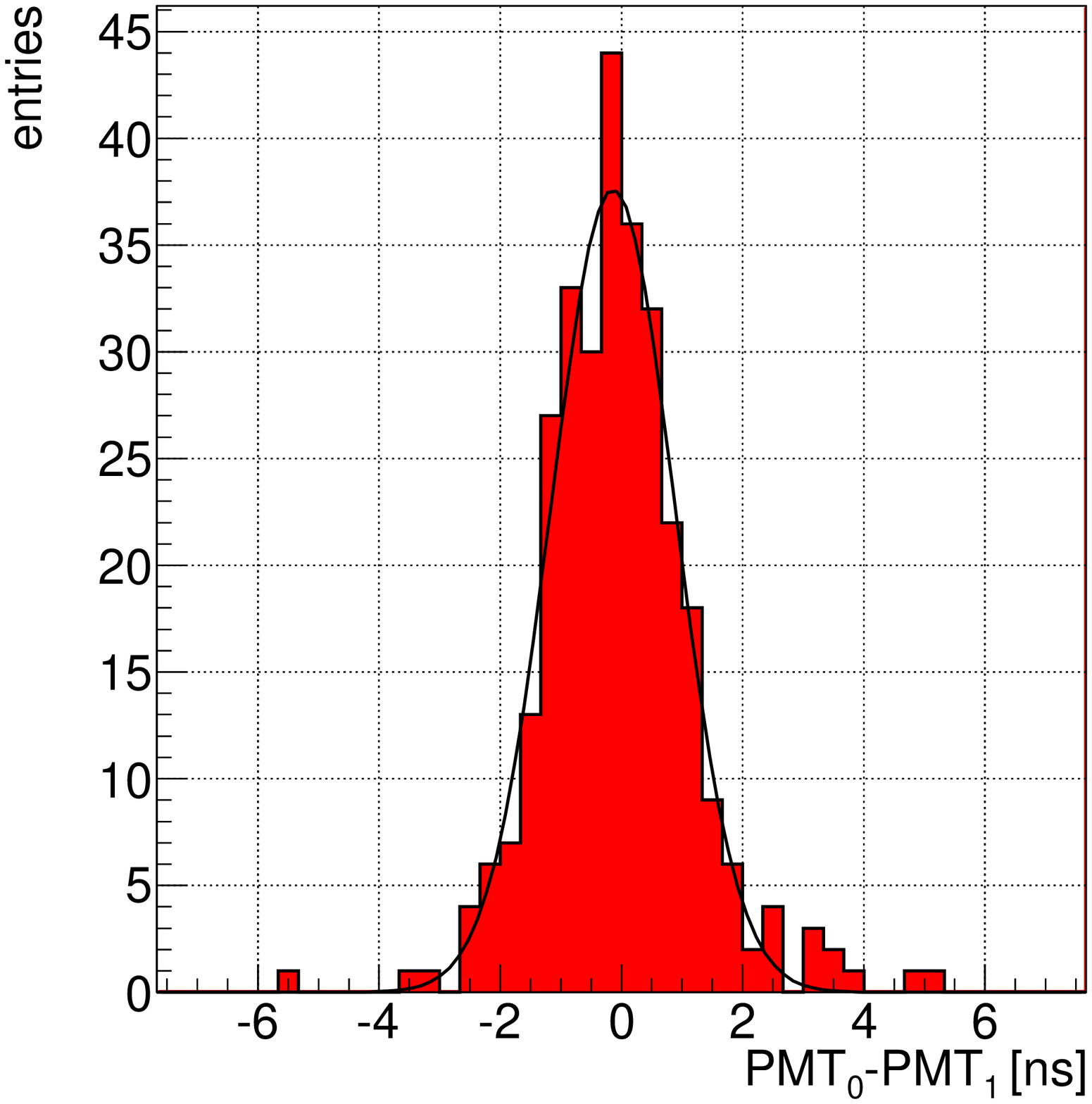}}\caption{\label{f-fig22_jap_pgaps}TDC time difference distribution of both PMTs attached to a paddle for a 2.5\,cm wide slice in the middle.}
\end{minipage}
\end{center}
\end{figure}

The next correction that needs to be applied is the timing offset between PMTs at opposite ends. Therefore, one of the PMTs gets assigned a constant time shift to ensure that the $t\sub{PMT,$0$}-t\sub{PMT,$1$}$ behavior as a function of the coordinate along the paddle is at 0 in the center of the paddle (Figure~\ref{f-fig21_jap_pgaps}). The data points were fitted with a straight line. The slope of the fit equals $2/v\sub{pulse}$ where $v\sub{pulse}$ is the effective pulse velocity inside the paddle from the hit position to the PMT. An average value of $v\sub{pulse}=(0.55\pm0.06)c$ over all paddles was measured and is about 20\,\% slower than the velocity naively calculated from the refractive index itself ($n=1.5$) because of photon reflections inside the scintillator material.

Gaussian fits to the $t\sub{PMT,$0$}-t\sub{PMT,$1$}$ distributions for the 2.5\,cm slice in the center of the paddles give an average width of $\sigma_t=(0.90\pm0.10)$\,ns that translates into an individual PMT timing resolution of $\sigma_t/\sqrt{2}=(0.64\pm0.07)$\,ns (Figure~\ref{f-fig22_jap_pgaps}). The timing resolution improvement using the outlined procedure is about 5\,\%. The TDC value measurement behavior can therefore be interpreted as stable and is not prone to big systematic corrections (Figure~\ref{f-fig20_jap_pgaps}).

\subsubsection{Velocity measurement\label{s-velo}}

For the actual velocity measurement, the individual paddle calibration of the last section has to be followed by the inter-paddle timing calibration. Therefore, a mean time for each paddle using both PMTs has to be calculated as well as the flight distance between them using the track fit. The flight distance between paddles with indices $j$ and $k$, $s_{j,k}$, is defined as: \be s_{j,k}=\sqrt{\sum_{i=1}^3(x_{i,j}-x_{i,k})^2}\ee with $x_i$ being the three-dimensional hit coordinates on the paddles. The mean time value for one paddle is calculated by \be\bar T=\frac12\left(t_\text{PMT,0}+t_\text{PMT,1}-\frac{l_\text{paddle}}{v_\text{pulse}}\right)\ee where $t_\text{PMT,0/1}$ are the corrected times of the PMTs connected to the same paddle, $l_\text{paddle}$ the paddle length, and $v_\text{pulse}$ the effective pulse velocity inside the paddle. The particle velocity $\beta$ is then calculated by \be\beta=\frac{s_{j,k}}{\left(\bar T_j-\bar T_k\right)\cdot c}.\ee As the trigger was set up to always include the TOF middle layer, the velocity calculation uses only combinations of top-to-middle and middle-to-bottom paddles. Furthermore, downward going particles are defined as having a positive $\beta$ value.

\begin{figure}
\begin{center}
\begin{minipage}[t]{.4\linewidth}
\centerline{\includegraphics[width=1.0\linewidth]{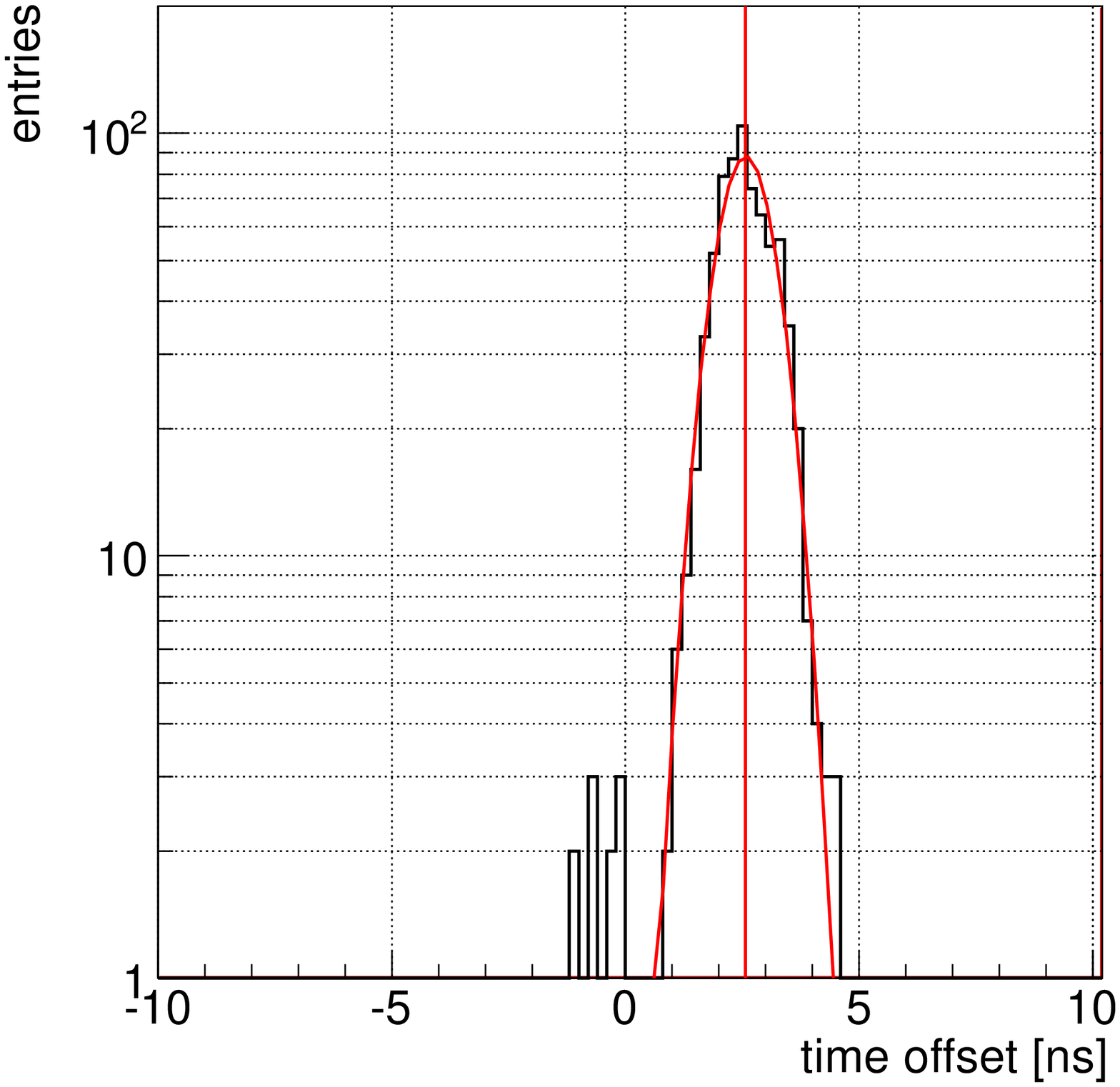}}\caption{\label{f-fig23_jap_pgaps}Paddle offset distribution for a paddle in the middle layer with respect to a paddle in the top layer for the choice of $\bar\beta=0.7$. The red curve denotes a Gaussian fit to the distribution and the vertical line the central value of this fit.}
\end{minipage}
\hspace{.1\linewidth}
\begin{minipage}[t]{.4\linewidth}
\centerline{\includegraphics[width=1.0\linewidth]{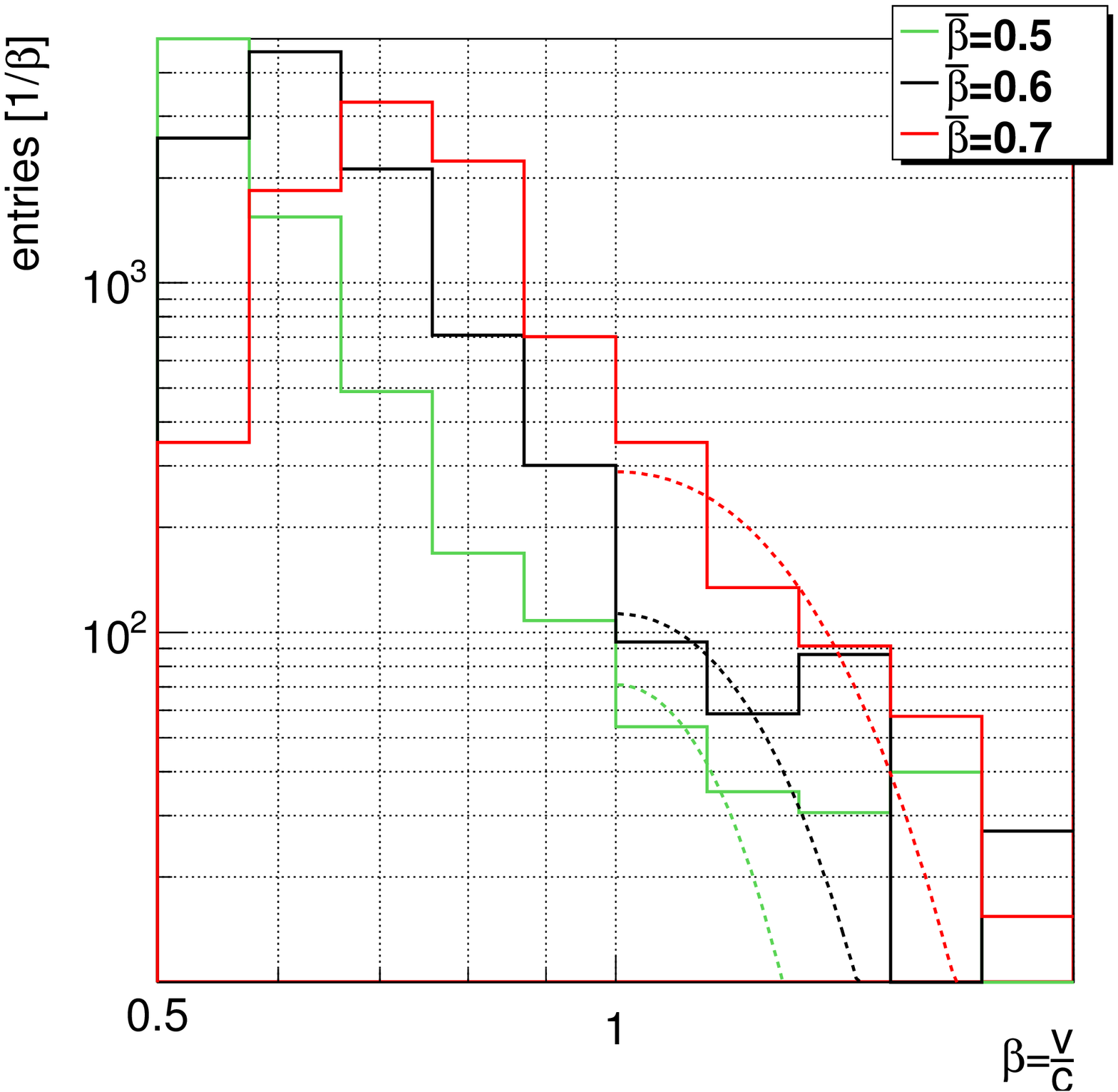}}\caption{\label{f-fig24_jap_pgaps}Histograms for different choices of $\bar\beta$ value (green:~0.5, black:~0.6, red:~0.7) for the paddle-to-paddle offset calculation with Gaussian fits of the distribution above $\beta=1$.}
\end{minipage}
\end{center}
\end{figure}

Constant time offsets between paddles in different layers distort the velocity measurement and need to be corrected for. If $\beta$ were known the timing offset $\Delta_{j,k}$ between two paddles $j$ and $k$ could be calculated by \be \Delta_{j,k} =\bar T_k-\bar T_j+\frac{s_{j,k}}{\beta\cdot c}\label{e-off}.\ee As there is no reason to assume that the $\beta$ distributions for different paddle-to-paddle combinations should have different mean values, the offset finding algorithm takes the mean $\bar\beta$ value as a free input parameter. The choice of $\bar\beta$ is discussed after outlining the algorithm. Only TOF paddles on the track with both PMTs having energy depositions above $|Z|>0.3$ and TDC hits were used for the offset calibration. For each event, all offsets between the hit paddles and a prechosen reference paddle were calculated by using the input $\bar\beta$ value, the distance known from the points where the track penetrates the TOF paddles, and the mean paddle time values. These offsets were histogrammed per paddle where the center value of a Gaussian fit to the distribution reflects the mean offset value for the choice of $\bar\beta$ value and reference paddle (Figure~\ref{f-fig23_jap_pgaps}). To take into account all geometrically allowed combinations, all six paddles in the top layer were used subsequently as reference paddles. The resulting offsets were combined and averaged.

The velocity was determined by all possible paddle-to-paddle combinations per event. The root mean square error $\sigma_\beta$ of the different velocity measurements was used to justify the choice of $\bar\beta$ in the following way: As a result of the finite TOF timing resolution, events with $\beta>1$ are expected. Therefore, a Gaussian centered at $\beta=1$ with the average of the velocity root mean square error of all events $\bar\sigma_\beta$ as standard deviation should be able to explain the $\beta$ distribution above the speed of light. The $\bar\sigma_\beta$ value for each choice of $\bar\beta$ value was calculated by averaging over the individual $\sigma_\beta$ values per event. Figure~\ref{f-fig24_jap_pgaps} shows the velocity distributions for three different choices of $\bar\beta$ with corresponding Gaussian fits with widths of $\bar\sigma_{\beta}$ where the only free parameter of the fit was the normalization factor. These fits underestimate the number of events above the speed of light. However, a systematic uncertainty band for the choice of $\bar\beta$ from 0.5 to 0.7 will be used in the following to illustrate the effect on the velocity measurement (Section~\ref{s-charge}). To avoid the difficulty of the absolute velocity measurement calibration in the future, it is foreseen to calibrate the full GAPS experiment with coincident LED test pulses and with test beam measurements.

Furthermore, the redundant beta measurements produced an improvement (by a factor of 2) in the timing resolution over single beta measurements (to \tilde0.7\,ns).

\section{Flux analysis\label{s-analysis}}

\subsection{Detection efficiencies and correction factors\label{s-eff}}

\begin{figure}
\begin{center}
\begin{minipage}[t]{.4\linewidth}
\centerline{\includegraphics[width=1.0\linewidth]{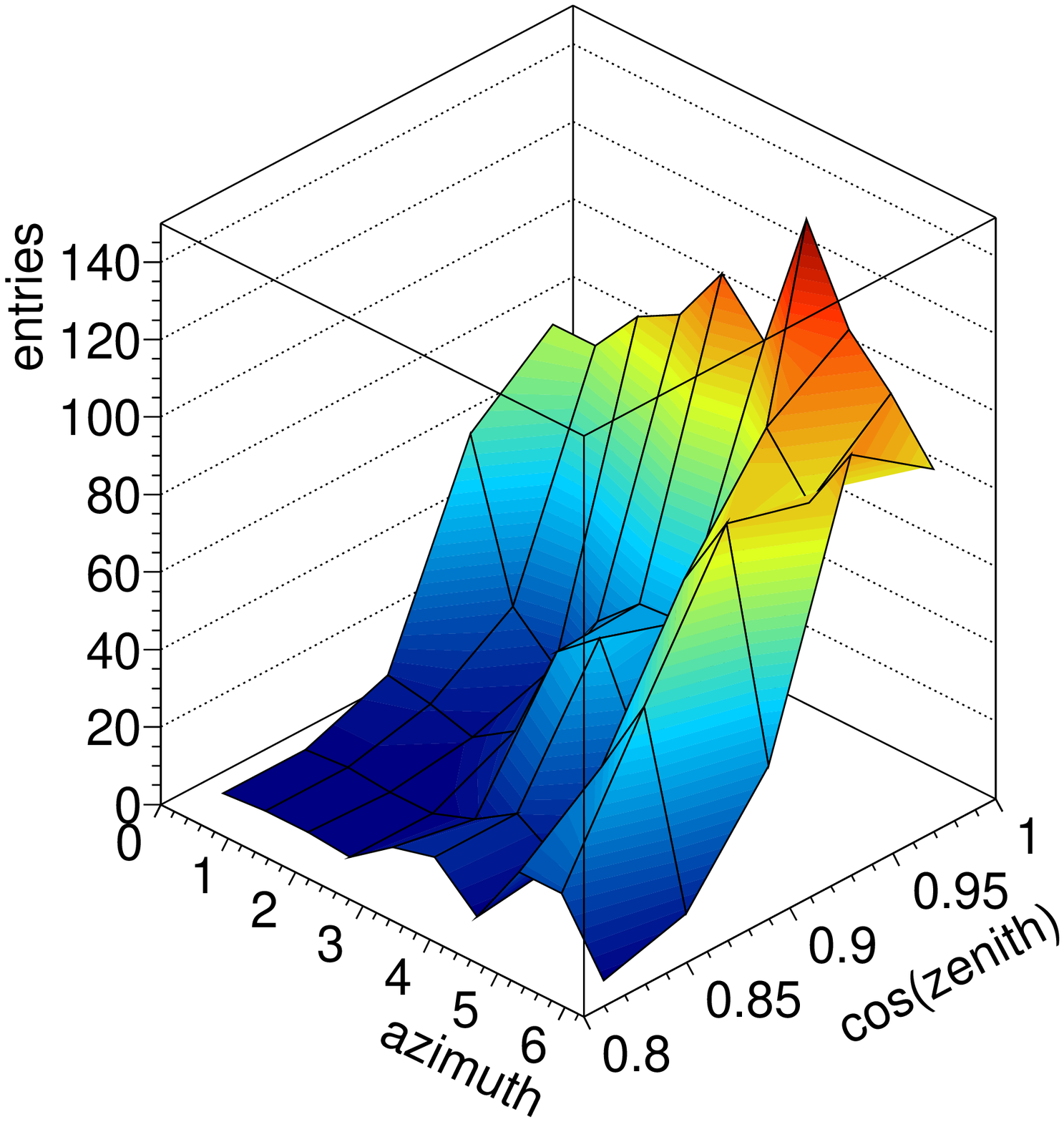}}\caption{\label{f-fig25_jap_pgaps}Directional distribution of clean events at float altitude.}
\end{minipage}
\hspace{.1\linewidth}
\begin{minipage}[t]{.4\linewidth}
\centerline{\includegraphics[width=1.0\linewidth]{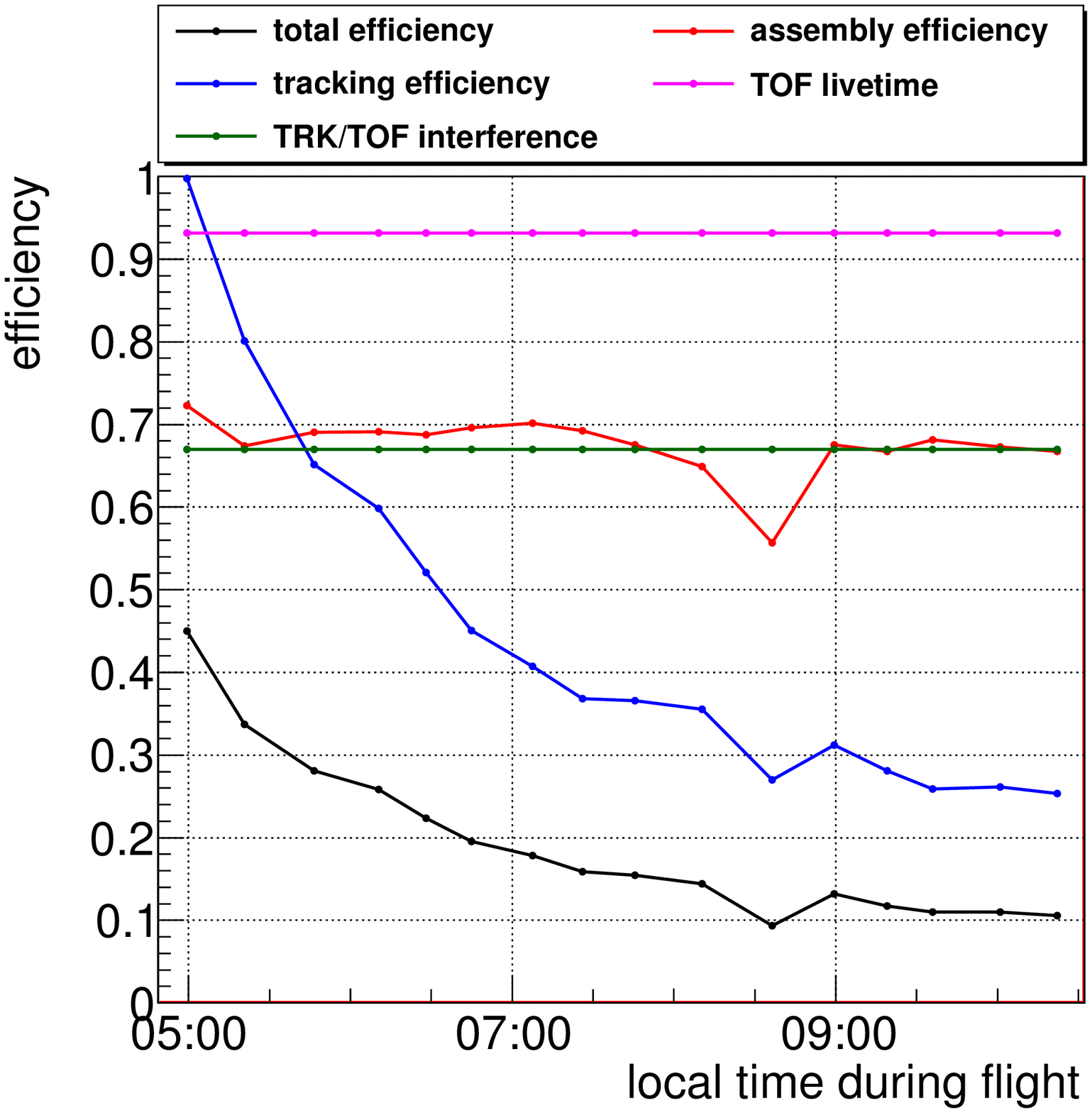}}\caption{\label{f-fig26_jap_pgaps}Summary of correction factors for flux analysis as a function of flight time. The total efficiency (black) is the result of multiplying the assembly efficiency (red), the tracking efficiency (blue), the relative TOF livetime (magenta), and the TRK/TOF interference factor (green).}
\end{minipage}
\end{center}
\end{figure}

The number of reconstructed events $N$ can be translated into a flux $F$ by taking into account  measurement livetime $L$ geometrical acceptance $A$, and detection efficiencies $\epsilon$:\be F=\frac{N}{L\cdot A}\cdot\frac1{\epsilon}.\ee The directional distribution of clean charged particle tracks as a function of $\cos\theta$ and the azimuth angle at float altitude is shown in Figure~\ref{f-fig25_jap_pgaps}. The visible structure can be explained by the locations of active strips and TOF paddles. Figure~\ref{f-fig26_jap_pgaps} summarizes the different efficiencies and correction factors that are needed to transform the pGAPS measurements into a non-instrument-specific flux and is explained in the following sections. The inverse of the black curve is the total systematic correction factor $1/\epsilon$ that needs to be applied to correct the measured number of clean events with at least two hit tracker strips to the number for the flux calculation.

\subsubsection{Event assembly efficiency}

In TOF trigger mode one TOF data packet and six individual data packets from the different tracker modules were issued at the same time. These data packets contained synchronized time counter values with 100\,ns resolution to be able to assemble individual subsystem data packets to full events during the analysis. The details of the event assembly algorithm are described in \ref{s-assembly}. Unfortunately, not all events could be assembled to complete events due to time counter readout failures. However, as each TOF trigger triggered the readout of the six tracker modules the event assembly efficiency can be defined as the number of complete events divided by the number of total individual tracker packets divided by six. The red curve in Figure~\ref{f-fig26_jap_pgaps} shows the behavior as a function of time with assembly efficiency values of about 70\,\% over the course of the flight. One dip shortly after reaching float altitude at 8:00\,am is visible and was caused by a flight computer reboot during the ascent from boomerang to float altitude and corresponding system adjustments. After this period the assembly efficiency showed the same value as before. The full GAPS payload will follow a different approach not only relying on synchronized time counter values for the event building, but on onboard event building electronics.

\subsubsection{TOF and tracker electronics interference}

An interference between the tracker and TOF electronics was present that manifested itself in a reduced rate when the tracker electronics were operated in TOF trigger mode. The effect was studied by looking at the TOF trigger rates when the tracker system was completely turned off and when the tracker system was fully operational in TOF trigger mode. This measurement was carried out during the launch preparations on ground and at the end of the flight and shows a constant value for these two very different environments of ($67\pm5$)\,\% (green curve). It was an unknown problem before the flight preparation in Taiki. Anyhow as mentioned above, the full GAPS readout system will follow a different approach that will make this issue obsolete.

\subsubsection{System livetime}

As the tracking system electronics had a significantly shorter deadtime than the TOF electronics (about a factor of 20) the relative livetime for the tracking system during TOF trigger mode was nearly 100\,\%. The TOF system showed a constant livetime of $(93.2\pm0.1)$\,\%, as discussed in Section~\ref{s-tof-live}.

\subsubsection{Acceptance\label{s-accep}}

The geometrical acceptance of the pGAPS detector using the TOF trigger condition (Section~\ref{s-trig}) and the clean track requirement (\ref{s-track}) was calculated by applying the Monte Carlo approach from \cite{sull} resulting in ($0.0116\pm0.0005$)\,m$^2$sr for particles coming from above and below, where the error on the acceptance reflects the uncertainty of the detector geometry. As previously discussed in Section~\ref{s-trk_stab}, the increasing temperature of the tracking system caused a change of geometric acceptance over the course of the flight and led to a decrease of particle tracks with at least two tracker hits. A correction factor can be estimated by studying the number of events with two tracker hits in comparison to the total number of events. The blue graph in Figure~\ref{f-fig26_jap_pgaps} shows this behavior normalized to the ratio of clean two tracker strip events over the number of total triggered events from the beginning of the flight when all detectors were active.

\subsection{Charged particle fluxes\label{s-charge}}

\begin{figure}
\centerline{\includegraphics[width=0.6\linewidth]{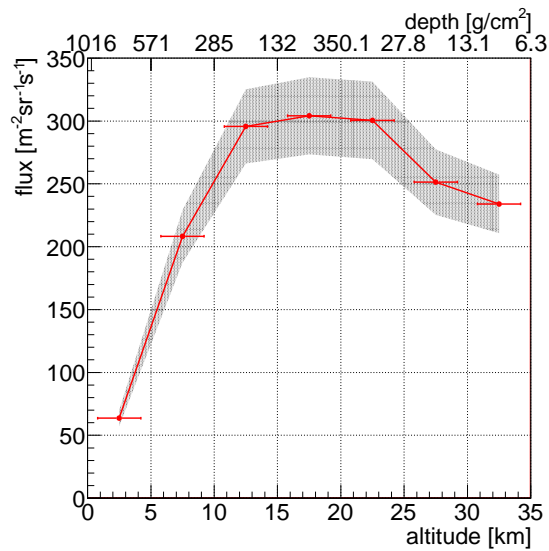}}\caption{\label{f-fig27_jap_pgaps}Total flux as a function of altitude. The gray shaded area depicts the systematic error from the flux correction explained in Section~\ref{s-eff}.}
\end{figure}

The charged particle flux as a function of altitude is shown in Figure~\ref{f-fig27_jap_pgaps}. The efficiencies discussed in the last sections were treated as systematic effects and are shown as an error band. At boomerang altitude (10--15\,km) the flux is about 30\,\% higher than at float altitude.

\begin{figure}
\centerline{\includegraphics[width=0.6\linewidth]{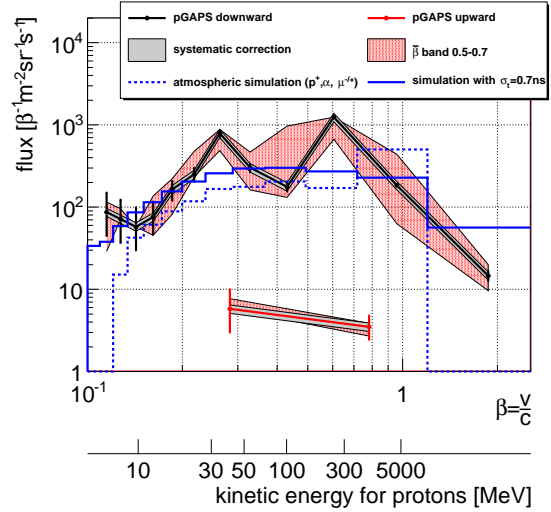}}
\caption{\label{f-fig28_jap_pgaps}Downward (black) and upward (red) going charged particle flux as a function of velocity for the period at float. The gray area depicts the systematic error from the flux correction and the red shaded area from the choice of $\bar\beta$ value  between 0.5 and 0.7. The dashed blue histogram shows the downward flux of protons, $\alpha$ particles, and muons at Taiki (33km) as predicted by air shower simulations \cite{planeto,phd,idm}. The solid blue histogram shows the results of the same simulations, but assuming 0.7\,ns TOF timing resolution. The additional $x$ axis illustrates the kinetic energy range assuming that all particles are protons.}
\end{figure}

Figure~\ref{f-fig28_jap_pgaps} shows the downward and upward fluxes measured at float altitude as a function of $\beta$. The bin widths were chosen to reflect the pGAPS TOF timing resolution. For a reliable $\beta$ value it was required that a paddle used for the calculation had clean energy deposition in both PMTs with $|Z|>0.3$ and non-zero timing values. The flux was scaled accordingly for events not fulfilling this quality cut. The additional $x$ axis illustrates the corresponding kinetic energy using the proton mass. In addition to the systematic error bands due to detection efficiencies (Section~\ref{s-eff}) and the choice of $\bar\beta$ value  (Section~\ref{s-velo}), the flux from air shower and geomagnetic simulations for Taiki at 33\,km altitude with PLANETOCOSMICS is shown \cite{planeto,phd,idm}. The dashed blue histogram shows the simulated combined proton, $\alpha$ particle, and muon flux assuming perfect timing resolution and exhibits a clear peak at $\beta=1$. Introducing a TOF timing resolution of $\sigma_t=0.7$\,ns and assuming an average distance between TOF paddles of 0.45\,m, entries from the sharp peak migrate to lower and higher velocity values (solid blue histogram). As shown in \cite{idm}, the atmospheric simulations are able to reproduce the fluxes in the atmosphere as measured by various experiments (e.g., BESS). However, these experiments did not report the atmospheric fluxes below kinetic energies of \tilde500\,MeV. The pGAPS measurements agree well with the simulations using 0.7\,ns TOF timing resolution for velocities above $\beta=0.7$, but show upward deviations at $\beta$ values of about 0.6 and 0.25 corresponding to kinetic energies for protons of \tilde250\,MeV and \tilde30\,MeV, respectively. Below $\beta=0.2$ the simulations describe the spectrum very well and also the range from 0.4 to 0.5 is in good agreement within the statistical and systematic errors. The low energy range is especially prone to atmospheric interaction and geomagnetic deflection physics effects 
and the low-velocity pGAPS results will be subject to further studies in the future.

The total measured flux is mostly composed of downward going particles and contains only a small fraction (\tilde1\,\%) of upward going particles. In this regard, it is important to mention that the upward going flux was not corrected for the components and material under the science part of the gondola (batteries, ballast hoppers, gondola bus, etc.). The upward coming flux is expected to be mostly composed of low-energy particles and therefore the material attenuates the actual flux significantly. 

\begin{figure}
\centerline{\includegraphics[width=0.6\linewidth]{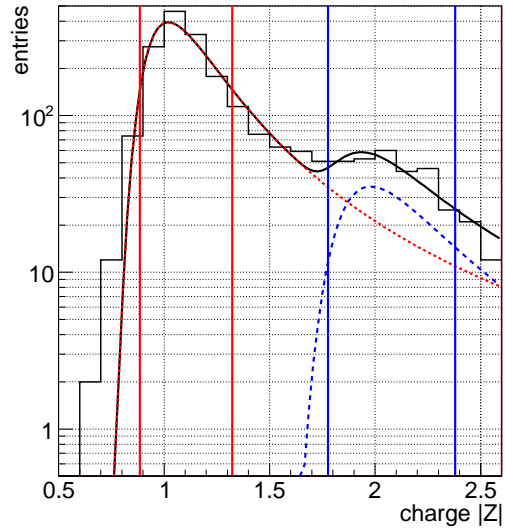}}
\caption{\label{f-fig29_jap_pgaps}Average charge $|Z|$ on track at float altitude. The black line shows a double peak Landau fit and the red (blue) lines show the individual contributions of the $|Z|=1$ ($|Z|=2$) distributions. The vertical lines indicate the 68.3\,\% confidence interval around the peaks.}
\end{figure}

The distribution of averaged energy depositions over all hit TOF and tracker detectors per event for the time at float reveals a clear peak at $|Z|=1$ and a second peak at $|Z|=2$ coming from $\alpha$ particles where the second peak was not observed on the ground (Figure~\ref{f-fig29_jap_pgaps}). The ratio of the integrals of the fitted Landau distributions for the $|Z|=2$ to $|Z|=1$ populations is $(10.4\pm0.7)$\,\%. For the same atmospheric depth of \tilde9\,g/cm$^2$, measurements by the BESS spectrometer for the 2001 flight from Ft. Sumner showed a ratio of the integral fluxes of $\alpha$ particles to protons and muons of about 14\,\% for the kinetic energy range of 0.5--10\,GeV$/n$ \cite{besspral}. The discrepancy can be explained by the geomagnetic cut-off, which is about twice as high at Taiki (\tilde8\,GV) than in Ft. Sumner (\tilde4\,GV) \cite{planeto,phd}. While protons with energies below the geomagnetic cut-off are abundantly produced in interactions of cosmic rays with the atmosphere, the probability of atmospheric production of $\alpha$ particles is very low. Therefore, the ratio of $\alpha$ particles to protons and muons is expected to drop with increasing geomagnetic cut-off. 

\subsection{Coincidence of charged particles and X-rays\label{s-chargex}}

The charged particle and X-ray results can be used to estimate the flux of events with coincident charged particles and atmospheric and cosmic X-rays in the range of antideuteron exotic atom ladder transitions. For the full payload, the geometric acceptance for charged particle tracks crossing all 10 tracker planes is about 3\,m$^2$sr. pGAPS measured an integrated charged particle flux of about 250\,m$^{-2}$sr$^{-1}$s$^{-1}$ at 33\,km. Assuming a rather pessimistic X-ray energy resolution, the interesting ranges for the antideuteron exotic atom ladder transitions are 27--33\,keV, 41--47\,keV, and 64--70\,keV. The integrated X-ray flux of these ranges was measured to be $29.3$\,m$^{-2}$sr$^{-1}$s$^{-1}$ (Fig.\ref{f-fig14_jap_pgaps}). The X-rays have to be reasonably close to the charged particle track. Therefore, this calculation assumes that the X-ray energy depositions have to occur in the detectors on the track, which translates into an X-ray acceptance of 0.44\,m$^2$sr for the coincidence calculation. The exotic atom ladder transitions will happen within 50\,ns. As a result, the flux of charged particles through the whole experiment in coincidence with one X-ray in the antideuteron exotic atom ladder transition range is $1.6\cdot10^{-4}$\,m$^{-2}$sr$^{-1}$s$^{-1}$ and with two X-rays $10^{-10}$\,m$^{-2}$sr$^{-1}$s$^{-1}$, respectively. These background fluxes should be taken as conservative upper limits, as they are not taking into account other particle identification techniques like penetration depth in the tracker, the development of the energy loss throughout the detector, and for the case of two X-rays that they must come from two different ladder transitions. In conclusion, requiring more than one X-ray in the right energy ranges alone suppresses this type of background extremely well.

\section{Conclusions}

The identification of dark matter is one of the most striking problems in physics and a low-energy cosmic ray antideuteron search has great potential in revealing deep insights. The GAPS experiment is specifically designed to perform this task. A prototype GAPS was successfully flown in June 2012. The purpose of this flight was to test and thoroughly analyze the concepts that form the basis for future flights. All goals for the flight were met and it was shown that the Si(Li) tracker detector modules and TOF worked reliably under flight conditions, the thermal model was verified \cite{isaacpaper}, and background particle and X-ray fluxes were measured. The detailed design work for the full GAPS payload has been started already.

\section*{Acknowledgments}

We thank G. Tajiri and D. Stefanik for the mechanical engineering support, and also thank J. Hoberman and B. Mochizuki for the development of the GAPS electronics. Furthermore, we would like to thank C. Hailey, K. Kamdin, P. Kaplan, M. Lopez-Thibodeaux, and T. Zhang for their contributions to the project. We thank the Scientific Balloon Office of ISAS/JAXA for the professional support of the pGAPS flight. This work is partly supported in the US by NASA APRA Grants (NNX09AC13G, NNX09AC16G) and the UCLA Division of Physical Sciences and in Japan by MEXT grants KAKENHI (22340073). K. Perez's work is supported by the National Science Foundation under Award No. 1202958.

\appendix

\section{Event reconstruction and track fit}

\subsection{Event assembly\label{s-assembly}}

\begin{figure}
\begin{center}
\begin{minipage}[t]{.4\linewidth}
\centerline{\includegraphics[width=1.0\linewidth]{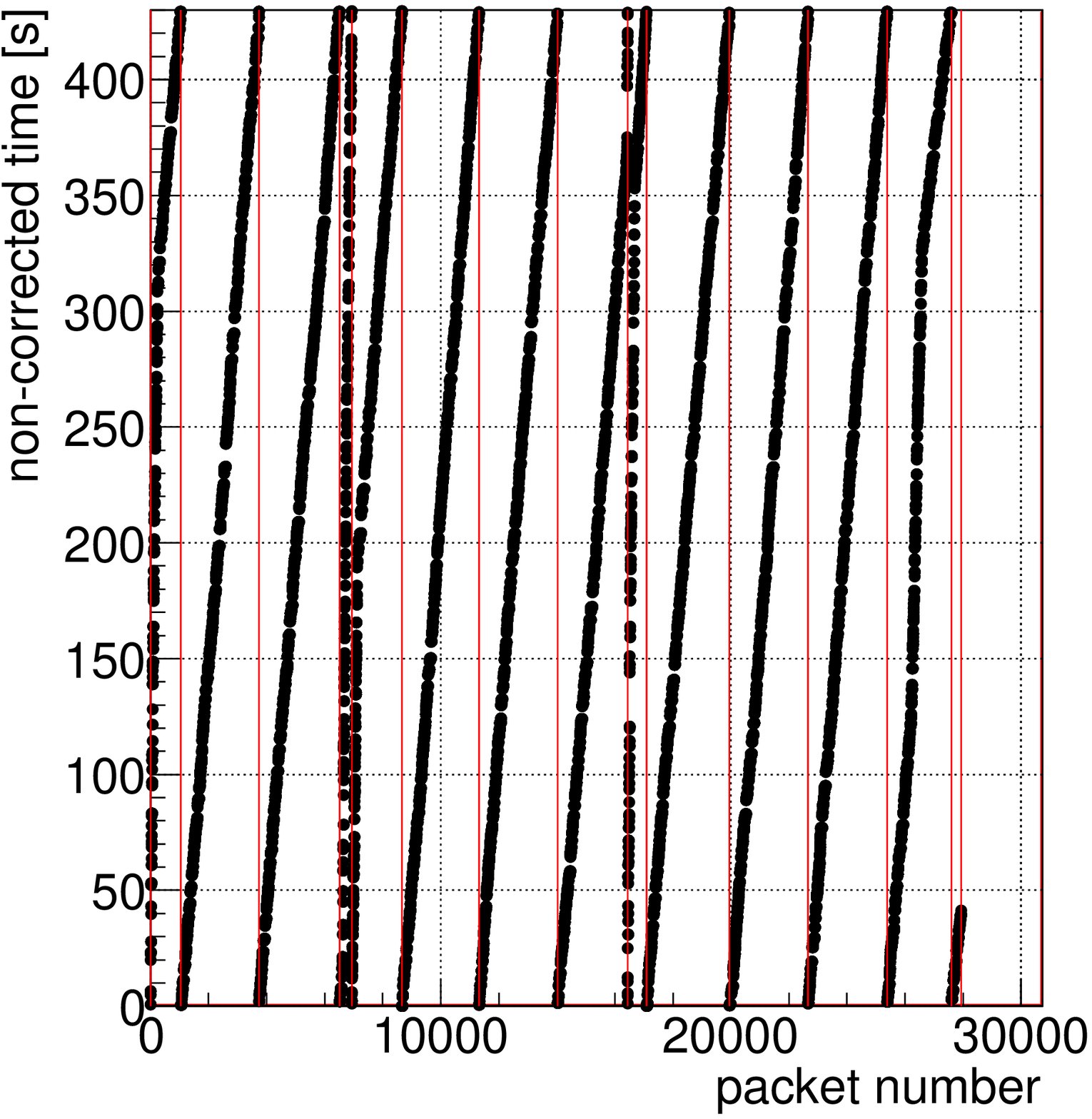}}\caption{\label{f-fig30_jap_pgaps}Uncorrected time counter values from the TOF system for a data excerpt from the end of the flight. Rollovers are marked with red vertical lines.}
\end{minipage}
\hspace{.1\linewidth}
\begin{minipage}[t]{.4\linewidth}
\centerline{\includegraphics[width=1.0\linewidth]{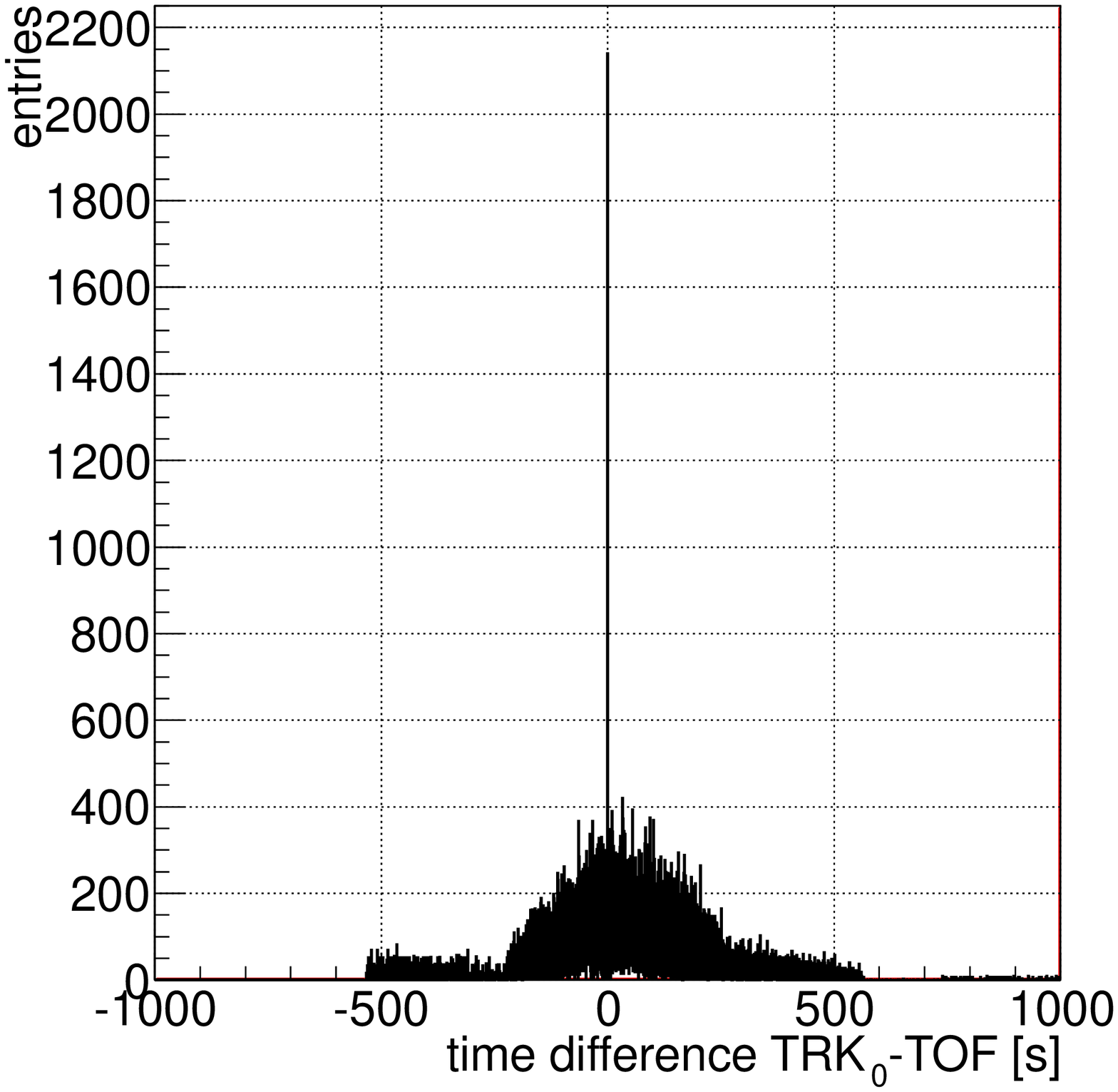}}\caption{\label{f-fig31_jap_pgaps}Initial time counter offset search with coarse binning.}
\end{minipage}
\end{center}
\end{figure}

The TOF and the two tracker electronics signal processing units were connected to the clock board on the flight computer. The clock board generated a continuous rectangular 10\,MHz pulse, which was connected to discriminator circuits in the TOF and tracker units and was used to count the number of transitions. The corresponding counters could be reset with a short individual pulse. The counter value was stored with 32\,bit precision and included in every event packet. In this way synchronized time counter values with 100\,ns resolution could be achieved between the different subsystem electronics. The 32\,bit counter was rolling over about every 7:15\,min and the first step in assembling full events of both tracker electronic units and the TOF was to find the rollover positions. Figure~\ref{f-fig30_jap_pgaps} shows the time counter values from the TOF system for a data excerpt from the end of the flight as a function of incoming packet order. The rollovers are clearly visible and the vertical lines mark the positions found by the analysis algorithm.

\begin{figure}
\begin{center}
\begin{minipage}[t]{.4\linewidth}
\centerline{\includegraphics[width=1.0\linewidth]{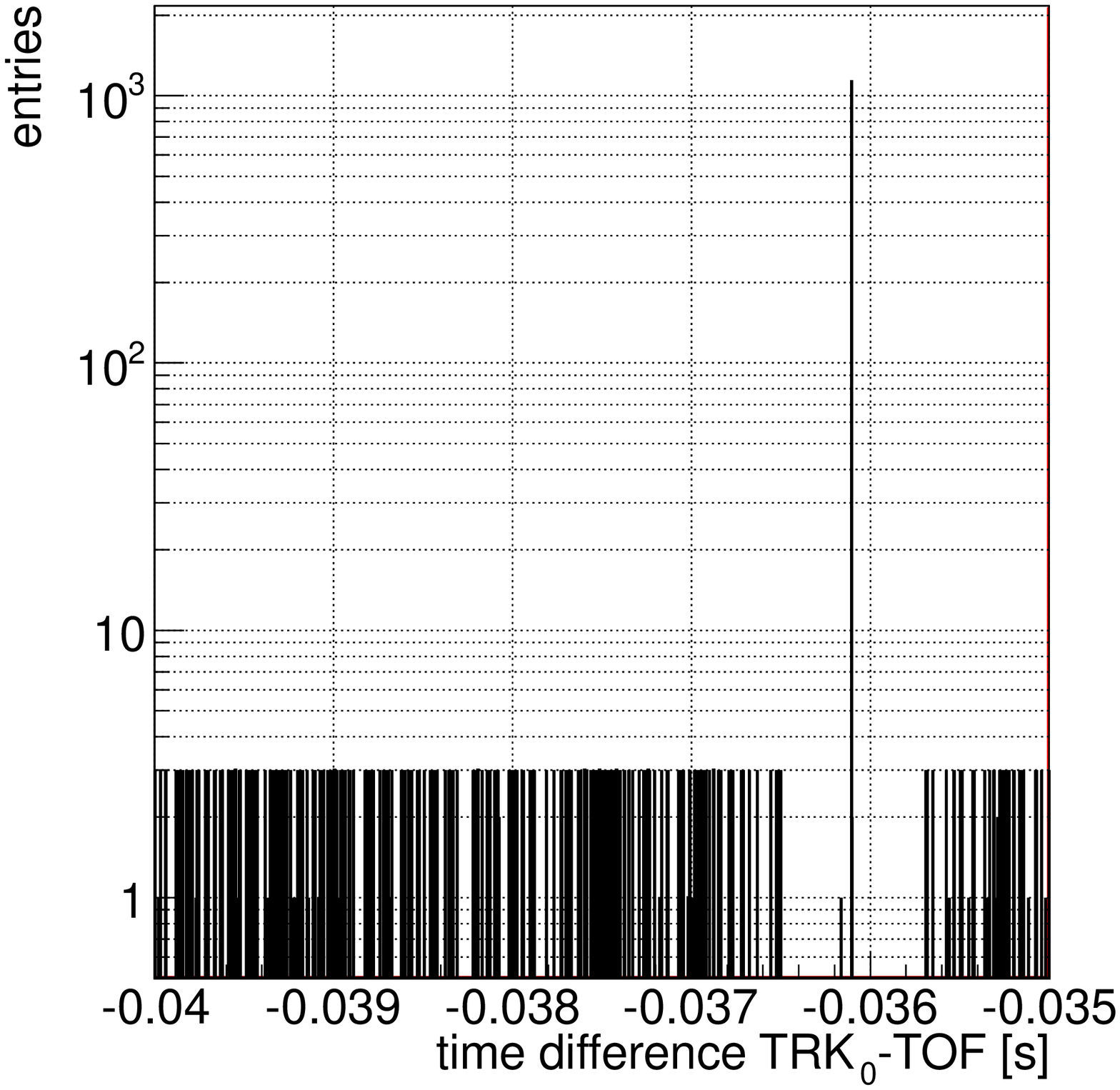}}\caption{\label{f-fig32_jap_pgaps}Time counter offset search inside the peak of Figure~\ref{f-fig31_jap_pgaps} with 100\,ns binning.}
\end{minipage}
\hspace{.1\linewidth}
\begin{minipage}[t]{.4\linewidth}
\centerline{\includegraphics[width=1.0\linewidth]{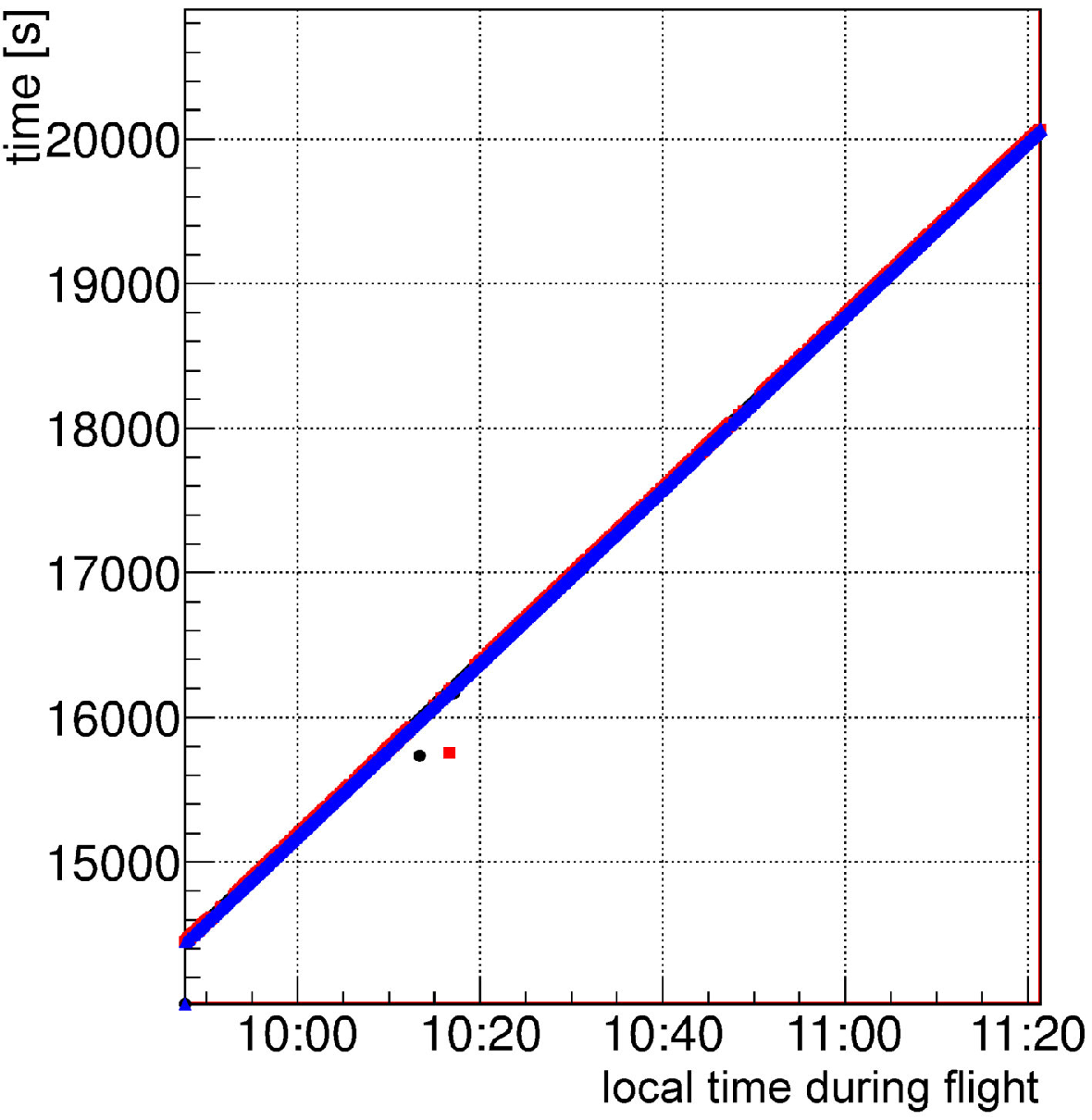}}\caption{\label{f-fig33_jap_pgaps}Fully corrected time counter values from the two tracker electronics units and the TOF system for the same part of the flight as in Figure~\ref{f-fig30_jap_pgaps}.}
\end{minipage}
\end{center}
\end{figure}

After adjusting for the number of rollovers by adding the corresponding number of seconds, the different electronics subsystems can be compared with each other. To find possible timing offsets between the subsystems, all rollover corrected time counter values of one subsystem were subtracted from the time counter values of another subsystem. The histogrammed differences are illustrated in Figure~\ref{f-fig31_jap_pgaps}. As expected from a constant clock offset a very sharp peak is visible. In the first step the histogram spanned a wide time range of 2000\,s to also allow for finding missed rollovers. In the second iteration the same differences were filled into a histogram only as wide as the maximum bin of the coarse histogram before with a 100\,ns bin width (Figure~\ref{f-fig32_jap_pgaps}). A very sharp peak covering only one bin is visible and is used as the timing offset between the two subsystems under study. This procedure was carried out for each data taking block between clock resets and sampled over time to study if the timing offsets were constant. It was found that the offsets stayed absolutely constant after a clock reset, but could change after clock counter resets. Figure~\ref{f-fig33_jap_pgaps} shows the result of the timing and offset calibration for all three electronics subsystems where the corrected time counter values for event data containing non-zero information follow a straight line compared to the also recorded UNIX computer clock time. Data packets were merged into one event if the time counter values had a difference smaller than 1\,\textmu s.

\subsection{Track fit\label{s-track}}

\begin{figure}
\centerline{\includegraphics[height=0.95\textheight]{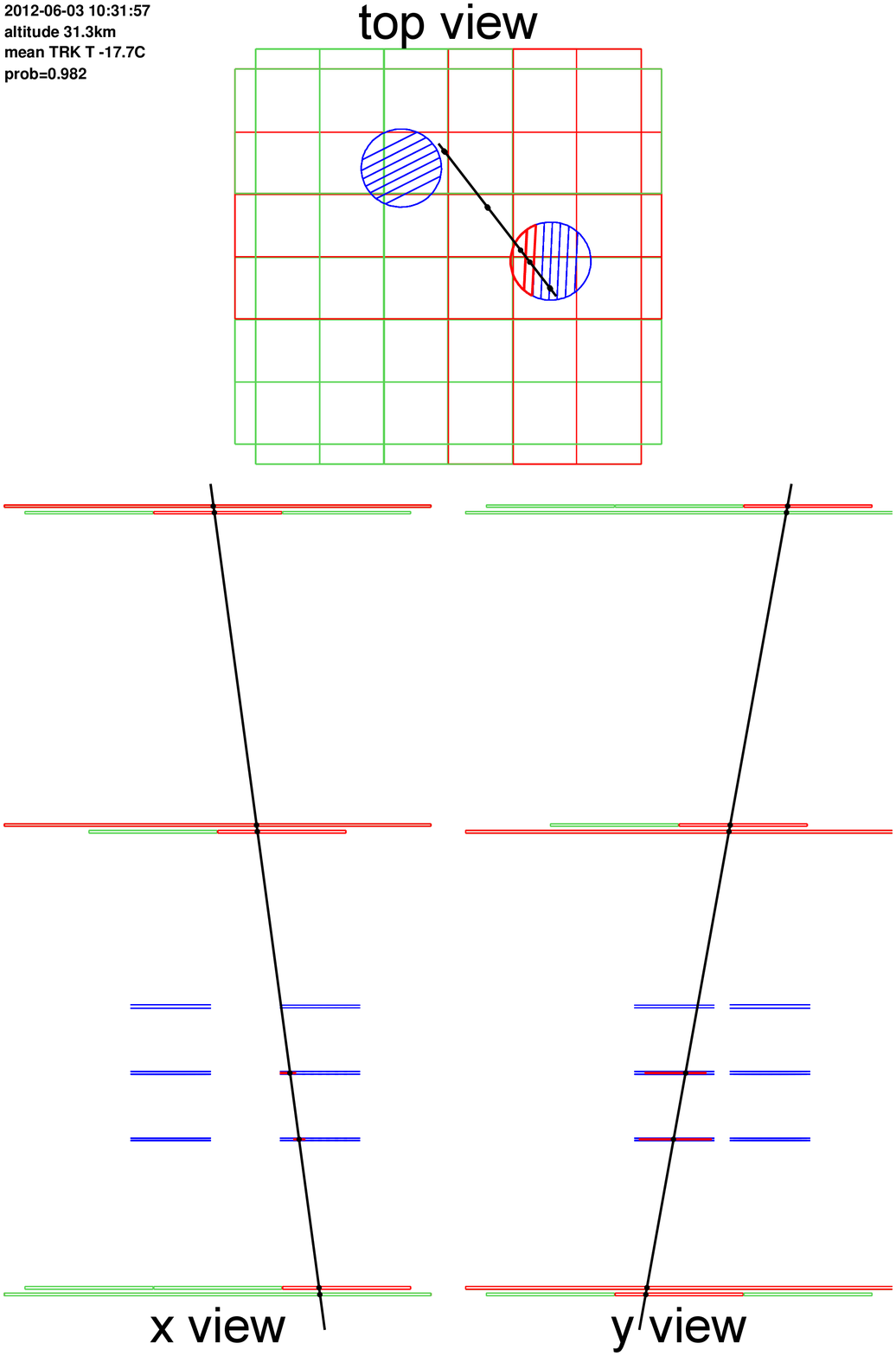}}
\caption{\label{f-fig34_jap_pgaps}Typical clean event.}
\end{figure}

The track fitting procedure has the goal to fit a parametrized straight line to charged particle tracks inside the detector using the least squares $\chi^2$ approach. A typical reconstructed event using the procedure described in the following is shown in Figure~\ref{f-fig34_jap_pgaps}. The TOF paddles ran along the $x$ and $y$ directions of the instrument coordinate system. The Si(Li) tracker is made up of two stacks with three circular modules each with a spacing of 8\,cm between the layers. The centers of the circular Si(Li) modules were offset from the central $z$ axis by 12\,cm and one stack was rotated by $-2.6^{\circ}$ and one by $117.4^{\circ}$ around the $z$ axis. The internal coordinate system of the Si(Li) modules is described by the cartesian coordinates $\vec u =(u,v,w)$ and needs to be rotated into the $\vec x= (x,y,z)$ absolute instrument coordinate system:

\be
\vec x = \mathcal R\cdot\vec u\quad\text{with}\quad\mathcal R =
\left(\begin{matrix} 
\cos\alpha_{0,1} 	& -\sin\alpha_{0,1}	& 0 \\ 
\sin\alpha_{0,1} 	& \cos\alpha_{0,1} 	& 0 \\ 
0	 	& 0		& 1
\end{matrix}\right),\\
\ee
where $\alpha_{0,1}$ are the rotation angles of the two stacks around the $z$ axis. Also the internal Si(Li) detector covariance matrix $\mathcal V$ needs to be transformed to the absolute coordinate system and can be obtained from standard error propagation:

\be
\mathcal U = \mathcal A\cdot\mathcal V\cdot\mathcal A^T\quad\text{with}\quad\mathcal V =
\left(\begin{matrix} 
\sigma_u^2	& 0		& 0 \\ 
0 		& \sigma_v^2 	& 0 \\ 
0	 	& 0		& \sigma_w^2
\end{matrix}\right)
\quad\text{and}\quad \mathcal A =
\left(\begin{matrix} 
\frac{\partial x}{\partial u} 	& \frac{\partial x}{\partial v} & \frac{\partial x}{\partial w} \\ 
\frac{\partial y}{\partial u} 	& \frac{\partial y}{\partial v} & \frac{\partial y}{\partial w} \\ 
\frac{\partial z}{\partial u} 	& \frac{\partial z}{\partial v} & \frac{\partial z}{\partial w}
\end{matrix}\right).
\ee

The $\sigma_{u,v,w}$ describe the errors in the internal coordinate system. The $\chi^2$ that needs to be minimized throughout the track fit is:
\be
\chi^2=\sum_i\chi^2_i=\sum_i \vec\Delta_i^T\mathcal U^{-1} \vec\Delta_i,
\ee
where $\vec\Delta_i$ denotes the difference vector between the position calculated from the straight line parametrization to the actual track point $i$ used for the fit. The $\chi^2_i$ calculation for the TOF points being part of the track fit used the same approach, but with a rotational angle $\alpha\sub{TOF}=0$, which simplifies the calculation and removes the non-diagonal elements of the covariance matrix.

\begin{figure}
\begin{center}
\begin{minipage}[t]{.4\linewidth}
\centerline{\includegraphics[width=1.0\linewidth]{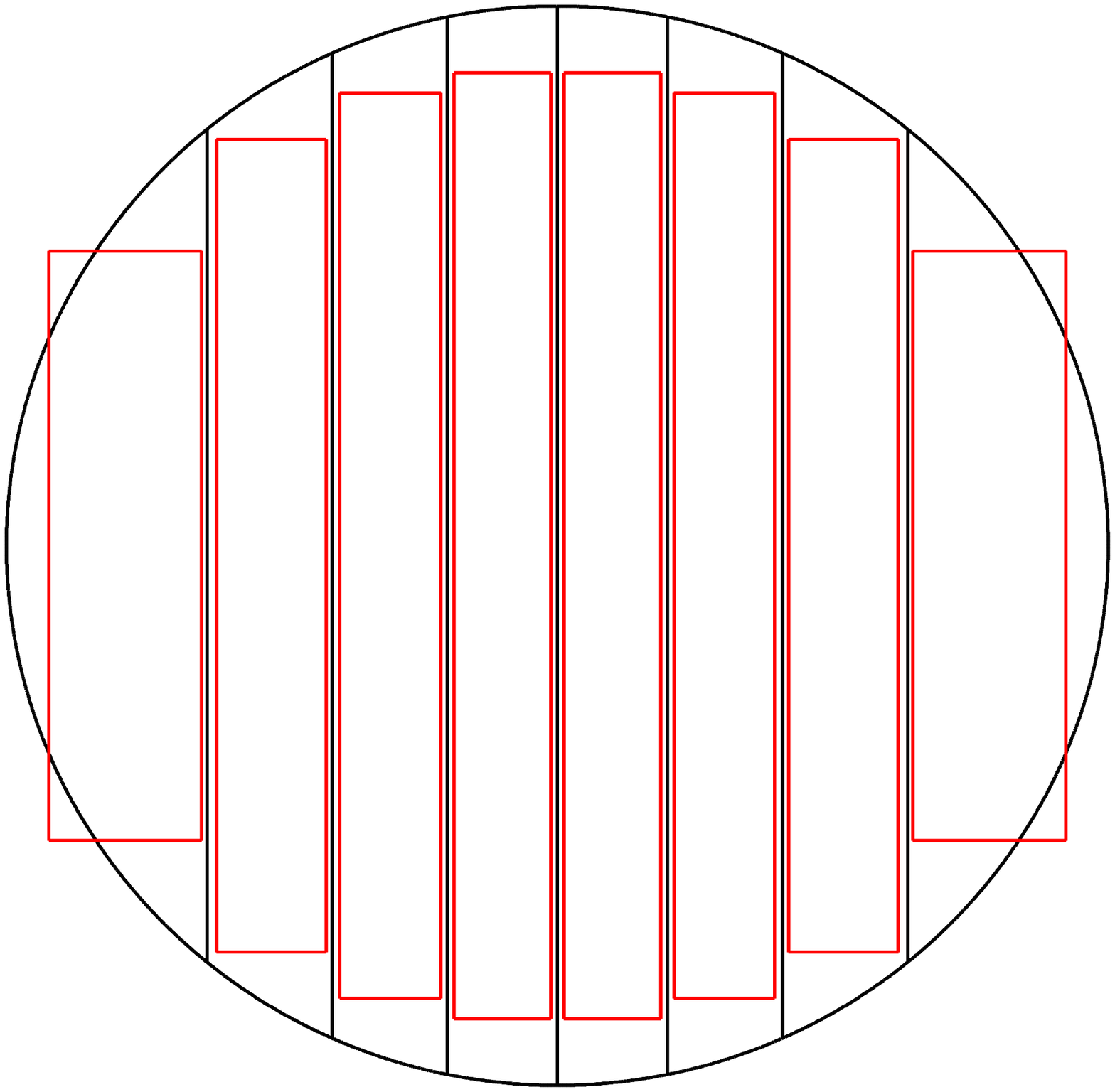}}\caption{\label{f-fig35_jap_pgaps}Schematic view of a circular Si(Li) detector where black vertical lines mark the grooves between different strips. The red boxes denote the errors used for the fit in the module plane.}
\end{minipage}
\hspace{.1\linewidth}
\begin{minipage}[t]{.4\linewidth}
\centerline{\includegraphics[width=1.0\linewidth]{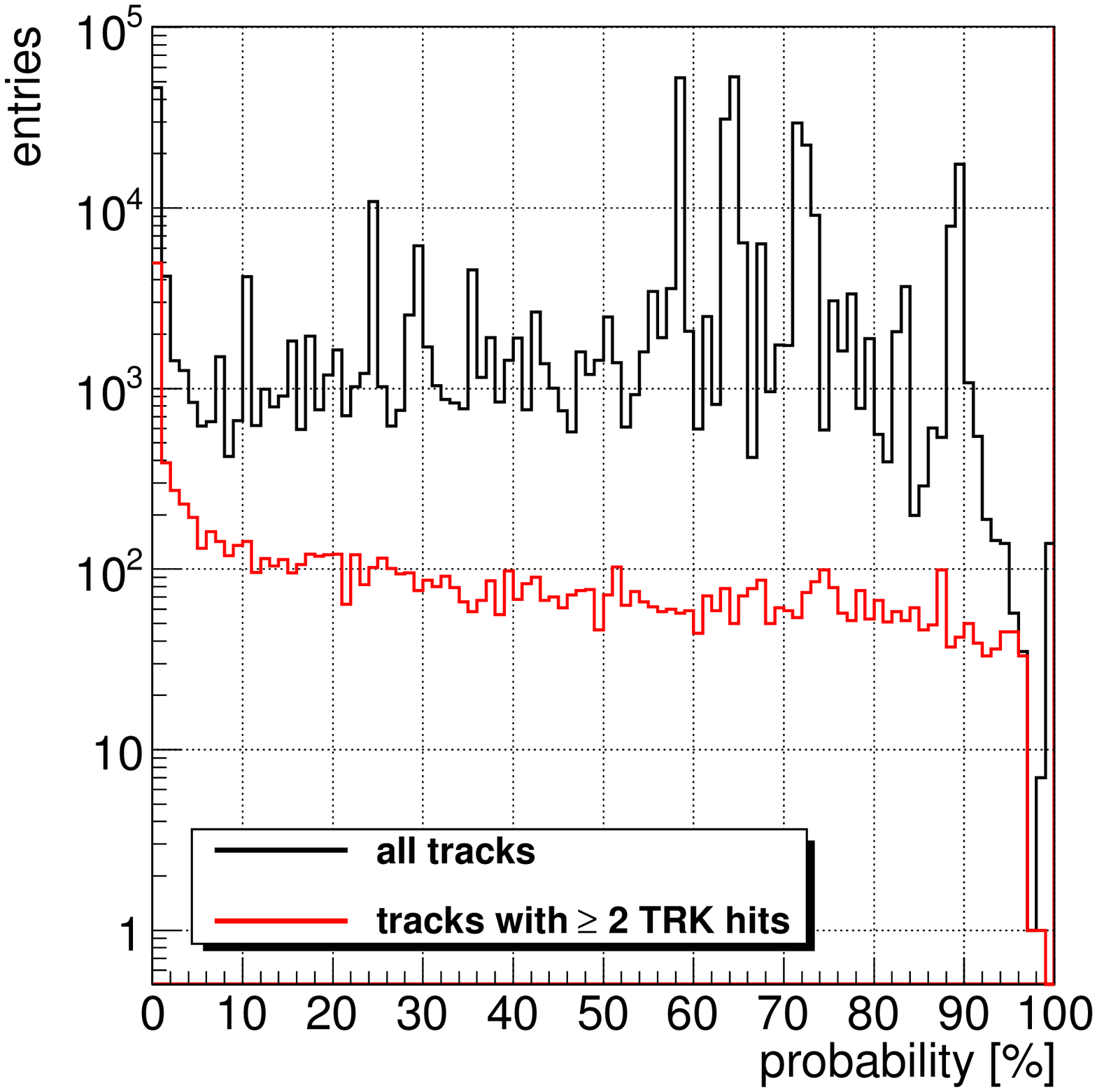}}\caption{\label{f-fig36_jap_pgaps}$p$-value distribution for requiring at least three hits of any type of TOF or tracker hit (black) and for requiring in addition at least two tracker hits being part of the event (red).}
\end{minipage}
\end{center}
\end{figure}

An active tracker strip for the track fit was defined as having an energy deposition above $|Z|>0.7$ while a TOF paddle was used for the track fit if the energy deposition of at least one PMT of the paddle was above $|Z|>0.3$ and also showed a non-zero timing value. Each active detector needed to be associated with a three-dimensional track coordinate and corresponding error bars. The center of gravity of a TOF paddle or Si(Li) detector strip was used as the average track coordinate for the particular detector volume. As an equivalent to the $1\sigma$ range used for fits with error bars in only one dimension, the three-dimensional error bars for a track point were chosen such that they span a volume of 68.3\% the size of the paddle or strip volume around the track point. The case for the rectangular box shaped TOF paddles is trivial and the error bars are set to be $\sqrt[3]{68.3\,\%}/2$ times the length of the paddle in its $x,y,z$ direction. The error boxes used for the case of the Si(Li) modules are shown in Figure~\ref{f-fig35_jap_pgaps}. The track fits were performed using the MINUIT fitting routines provided by ROOT \cite{minuit,root}.

A track is considered to be good if the requirement $p\geq0.05$ was fulfilled, where the $p$-value is defined as:\be p=\int_{\chi^2}^\infty f(t,n)\text{d}t\ee with the number of degrees of freedom $n$ and the $\chi^2$ probability density function $f(\chi^2,n)$. $n$ is the number of points used for the track fit subtracted by two. From purely statistical effects it is expected that the $p$-value distribution is uniform. A peak at $p=0$ corresponds to too many large $\chi^2$ values and is not in agreement with statistical fluctuations. For track quality reasons only the nearly uniform part above 5\,\% of the distribution was taken into account (Figure~\ref{f-fig36_jap_pgaps}). Clean tracks used for the analysis were required to have more than two hits in any active detector volume and in addition to show at least two tracker hits with energy depositions $|Z|>0.7$. Figure~\ref{f-fig36_jap_pgaps} demonstrates the power of this requirement as the tighter tracker constraint (red histogram) makes the distribution much cleaner than the looser definition (black histogram) because the TOF position resolution is much coarser than the tracker resolution.


\begin{thebibliography}{99}
\bibliographystyle{alpha}
\bibitem{darkmatter}
K. Freese, Europ. Astron. Soc. Pub. Ser. 36, 113 (2009).
\bibitem{pamela}
O. Adriani et al., Nature 458, 607 (2009).
\bibitem{pbarpamela}
O. Adriani et al., Phys. Rev. Lett., 105, 121101 (2010).
\bibitem{amsposi1}
M. Aguilar et al., Phys. Lett. B 646, 145 (2007).
\bibitem{amsposi2}
M. Aguilar et al., Phys. Rev. Let.. 110, 141102 (2013).

\bibitem{dbarsusy}
F. Donato et al., Phys. Rev. D 62, 043003 (2000).
\bibitem{dbarprod}
R. Duperray et al., Phys. Rev. D 71, 083013 (2005).
\bibitem{ibarra20132}
A. Ibarra and S. Wild., Phys. Rev. D 88, 023014 (2013).
\bibitem{forn}
N. Fornengo et al., J. Cosmol. Astropart. P. 09, 031 (2013).

\bibitem{dbarpbh}
A. Barrau et al., Astron. Astrophys. 398 (2), 403 (2003).
\bibitem{dbarpbh2}
A. Barrau et al., Phys. Rev. D 69, 105021 (2004).
\bibitem{dbarbaer}
H. Baer et al., J. Cosmol. Astropart. P. 12, 008 (2005).
\bibitem{dbarback}
F. Donato et al., Phys. Rev. D 78, 043506 (2008).
\bibitem{kadastik2010}
M. Kadastik et al., Phys. Lett. B 683 (4–5), 248 (2010).
\bibitem{antideuteroncui}
Y. Cui et al., J. High Energy Phys. 1011, 017 (2010).
\bibitem{gravi}
M. Grefe, J. Phys.: Conf. Ser. 375 012035 (2012).
\bibitem{ibarra20131}
A. Ibarra et al., J. Cosmol. Astropart. P. 1302, 021 (2013).

\bibitem{susydm}
G. Jungman et al., Phys. Rep. 267, 195 (1996).
\bibitem{kkdm}
D. Hooper et al., Phys. Rep. 453, 29 (2007).

\bibitem{bess}
H. Fuke et al., Phys. Rev. Lett. 95, 081101 (2005).

\bibitem{gaps}
C. Hailey,  New J. Phys. 11, 105022 (2009).
\bibitem{amsdbar}
V. Choutko et al., Proc. 30th Int. Cosmic ray Conf. 4, 765 (2007).
\bibitem{gapssens}
T. Aramaki, (to be published) (2013).
\bibitem{dbarsens}
P. von Doetinchem et al., Proc. 10th Symposium on Sources and Detection of Dark Matter and Dark Energy in the Universe (UCLA Dark Matter 2012) (2012), to be published.

\bibitem{solarcycle}
NASA, http://solarscience.msfc.nasa.gov/SunspotCycle.shtml (June 2013).

\bibitem{isaacpaper}
S. Mognet et al., Nucl. Instr. Meth. A 735, 24 (2014).

\bibitem{detdev}
T. Aramaki et al., Nucl. Instr. Meth. A 682, 90 (2012).

\bibitem{nct05}
W. Coburn et al., Proc. SPIE 5898, 589802 (2005).

\bibitem{taiki}
H. Fuke, D. Akita, I. Iijima, et al., Adv. Space Res. 45, 490 (2010).
\bibitem{jaxbal}
J. Nishimura and H. Hirosawa, Adv. Space Res. 1, 239 (1981).

\bibitem{planeto}
L. Desorgher, http://cosray.unibe.ch/\tilde laurent/planetocosmics (June 2013).
\bibitem{phd}
P. von Doetinchem, PhD thesis, RWTH Aachen University, (2009).
\bibitem{idm}
P. von Doetinchem et al., PoS(IDM2010) 063, (2010).

\bibitem{sull}
J. Sullivan, Nucl. Instr. Meth. A 95, 5 (1971).

\bibitem{besspral}
K. Abe et al., Phys. Lett. B 564, 8 (2003).

\bibitem{minuit}
F. James and M. Roos, Comput. Phys. Commun. 10, 343 (1975).
\bibitem{root}
R. Brun and F. Rademakers, Nucl. Instr. Meth. A 389, 81 (1997).

\end{thebibliography}
\end{document}